\documentclass[12pt]{article}

\pdfoutput=1 

\usepackage{graphicx}% Include figure files
\usepackage{dcolumn}% Align table columns on decimal point
\usepackage{bm}% bold math
\usepackage{subcaption}

\usepackage{amsmath}
\usepackage{amssymb}
\usepackage{mathtools}
\usepackage{authblk}
\usepackage{xcolor}
\usepackage{soul}
\numberwithin{equation}{section}

\usepackage{cite} % For ranges in citations

\usepackage[
stretch=10, shrink=10, 
protrusion=true,
expansion=true
]{microtype}

\usepackage{multirow}
\newcommand{\rbsix}{RB\,6}
\newcommand{\rsbsix}{RSB\,6}
\newcommand{\sqtlb}{SQTLB}

\newcommand{\CphiD}{C^{\varphi D}}
\newcommand{\QphiD}{Q_{\varphi D}} 
\newcommand{\Cphibox}{C^{\varphi \Box}}
\newcommand{\Qphibox}{Q_{\varphi \Box}} 
\newcommand{\Cphi}{C^{\varphi}}
\newcommand{\Qphi}{Q_{\varphi}}

\newcommand{\Cphithree}{C^{\varphi}_3}

\newcommand{\yt}{y_t}
\newcommand{\Cphiboxthree}{C^{\varphi \Box}_3}
\newcommand{\CphiDthree}{C^{\varphi D}_3}

\newcommand{\nmSq}{n_{m^2}}
\newcommand{\nlambda}{n_{\lambda}}
\newcommand{\nC}{n_{C}}
\newcommand{\nphi}{n_{\phi}}

\newcommand{\bea}{\begin{eqnarray}}
\newcommand{\eea}{\end{eqnarray}}

\usepackage{xspace}
\newcommand{\GeV}{\ensuremath{\text{Ge\kern-0.15ex V}}\xspace}
\newcommand{\TeV}{\ensuremath{\text{Te\kern-0.1ex V}}\xspace}

\newcommand{\Ordo}[1]{\mathcal{O}{\left(#1\right)}}
\newcommand{\diff}[1]{\mathrm{d}{#1}~}

\usepackage{csquotes} %smart quotes
  \MakeOuterQuote{"} %for automatic inverted commas
\usepackage{lmodern} %Updated version of the computer modern font
    \usepackage[
    margin=1.25in,
    ]{geometry}
\usepackage{xpatch}
\xpretocmd{\eqref}{Eq.~}{}{}

\usepackage{hyperref} 
\usepackage{cleveref}
\hypersetup{colorlinks=true,allcolors=blue,linktocpage=true,breaklinks=true}

\title{\bf A Catalog of First-Order\\ Electroweak Phase Transitions in the Standard Model Effective Field Theory}
\author{Eliel Camargo-Molina\thanks{eliel.camargo-molina@physics.uu.se}}
\author{Rikard Enberg\thanks{rikard.enberg@physics.uu.se}}
\author{Johan Löfgren\thanks{johan.lofgren@physics.uu.se}}
\affil{Department of Physics and Astronomy,\\ Uppsala University,\\ Box 516, SE-751 20 Uppsala, Sweden}

\date{June 27, 2025}% It is always \today, today,
\begin{document}

\maketitle

%================================================================================================================
\begin{abstract}
%================================================================================================================
\noindent{}We use modern dimensionally-reduced effective field theory methods, with careful attention to scale hierarchies, to analyze and catalog the types of first-order electroweak phase transitions that are possible in the Standard Model Effective Field Theory (SMEFT). Our calculations lay the necessary groundwork to perform gauge invariant, properly resummed perturbative expansions, and therefore address many of the theoretical problems with phase transition calculations.

We find three types of configurations of the scalar potential that allow for a first-order phase transition, namely tree-level barriers, radiative barriers, or radiative symmetry breaking through the Coleman-Weinberg mechanism. We also find versions of these with significant supercooling.
  
We perform a global likelihood scan over the Wilson coefficients of SMEFT operators involving only the Higgs field, to identify parameter regions that exhibit these first-order phase transitions and are consistent with experimental and theoretical constraints.

We comment on the possibilities for electroweak baryogenesis within the SMEFT, and roughly estimate if the gravitational wave spectra generated by the phase transitions are detectable.

\end{abstract}

\tableofcontents

%================================================================================================================
\section{Introduction}
%================================================================================================================
The question of whether the electroweak phase transition (EWPT) was an abrupt first-order phase transition or a continuous transition is intimately tied to the Higgs sector. An abrupt transition would occur via bubble nucleation and could therefore explain the observed matter-antimatter asymmetry of the Universe via the mechanism of electroweak baryogenesis~\cite{Morrissey:2012db}, making it important to understand what types of Higgs sectors can generate such a transition, and whether any region of their parameter space is allowed by experimental constraints. Furthermore, such a phase transition could plausibly source gravitational waves that can be detected by future space-based gravitational wave interferometry experiments such as DECIGO~\cite{Kawamura:2011zz}, LISA~\cite{Caprini:2015zlo,Caprini:2019egz}, TAIJI~\cite{Ruan:2018tsw}, or TIANQIN~\cite{TianQin:2015yph}. 

In this paper, we focus on the Standard Model Effective Field Theory (SMEFT)~\cite{Buchmuller:1985jz,Grzadkowski:2010es}, following Wheeler's principle of \textit{radical conservatism}~\cite{Thorne:2019scz}---that is, we take the SMEFT as a well-established BSM model and we push it in interesting, less established directions. The SMEFT is a bottom-up effective field theory (EFT) that contains all gauge invariant operators constructed from the Standard Model (SM) fields at a given mass dimension, in order to account for unknown heavy physics beyond the SM (BSM). The SMEFT is phenomenologically a very important model in BSM physics to consistently parameterize and search for new physics, and there are a large number of global fits and experimental searches restricting its parameters. It is also one of the cleanest BSM cases for the EWPT, since there is only one scalar field involved in the transition.

Our goal in this paper is twofold: 
\begin{enumerate}
\item
\textbf{Our first goal} is to construct a catalog of possible first-order phase transitions within the SMEFT, as an extension and refinement of our previous study in Ref.~\cite{Camargo-Molina:2021zgz}. The construction of the catalog itself provides a pedagogical exploration of first-order phase transitions using power-counting methods, and the catalog can support future studies of models featuring $\phi^6$ terms in the potential. We will furthermore use the catalog to map out for which regions of parameter space the SMEFT can support a first-order phase transition, and we will combine these constraints with collider data and other constraints to see what barriers are phenomenologically viable, following the principle that combinations of collider studies and EWPT conditions is one of the few promising ways to restrict the Higgs sector~\cite{Ramsey-Musolf:2019lsf}. In addition, we will make some preliminary estimates of the possibilities for detecting gravitational waves from the different types of transitions, and the viability of electroweak baryogenesis.

\item
\textbf{Our second goal} is to advocate for computing the properties of the phase transition using dimensionally reduced 3D Euclidean effective field theory methods, and to do this for a realistic, phenomenologically interesting BSM model. Too often, phenomenological studies of the EWPT use what we may call the ``4D method,'' which consists in using the one-loop effective potential with a Coleman-Weinberg (CW) term and a finite-temperature term, and performing a ring resummation (also called daisy resummation). It is now well-known that this method gives gauge-dependent results and very large theoretical uncertainties, see, for example, \cite{Lofgren:2023sep} and other works cited below.
\end{enumerate}
Let us now discuss these two goals.

For a first-order phase transition to be possible within the SMEFT there needs to be a barrier in the effective potential for the Higgs field. We include SMEFT operators up to dimension-six, of which there are three operators involving only the Higgs field and derivatives. Only one of these enters the tree-level scalar potential; the other two  contribute to the Higgs mass and will be included in our calculations. 

This means that we need to find in what possible ways a barrier between two minima can be generated in a gauge theory with a $\phi^6$ term in the potential,
\begin{equation}
    V(\phi)=-\frac{1}{2} m^2 \phi^2+\frac{1}{8}\lambda \phi^4-\frac{1}{8}\Cphi\phi^6.
\end{equation}
We follow the conventions of~\cite{Dedes:2017zog} for the SMEFT operators, and we consider $\Cphi<0$ to ensure the potential is bounded from below.\footnote{Because we are considering an EFT we could in principle consider potentials unbounded from below and simply assume that there are higher dimension operators that make the potential stable. But considering such a case would prevent us from studying the details of spontaneous symmetry breaking, and would complicate the question of perturbativity as it would require even higher-dimensional operators to become important.}

There have been a number of previous studies of the electroweak phase transition in the SMEFT or in more restricted models with dimension-six operators. The first such study was in Ref.~\cite{Grojean:2004xa}, with later work in, e.g., Refs.~\cite{Ham:2004zs, Bodeker:2004ws,Huber:2007vva,Delaunay:2007wb,Cai:2017tmh,deVries:2017ncy,Chala:2018ari,Ellis:2018mja,Croon:2020cgk,Postma:2020toi,Hashino:2022ghd,Banerjee:2024qiu,Qin:2024dfp} mainly focusing on tree-level barriers arising from the dimension-six operator, which allows a negative Higgs quartic coupling $\lambda$. This, however, requires a low cutoff scale: it has been questioned whether such a low cutoff is reasonable~\cite{Damgaard:2015con, Postma:2020toi}, and it has been argued dimension eight operators also become necessary~\cite{Chala:2018ari}.

The last few years have seen a rapid development in the methods used to study first-order phase transitions, allowing us to consistently describe models in which a barrier can arise in different ways. The modern methods are based on EFT principles, which gives an intuitive physical picture and a built-in region of applicability~\cite{Gould:2021ccf}---each form of barrier considered here comes with its own perturbative description. There is ongoing interest and active research in refining perturbation theory for describing first-order phase transitions~\cite{Croon:2020cgk, Gould:2021oba, Schicho:2022wty,Ekstedt:2022ceo, Athron:2022jyi, Lewicki:2024xan, Chala:2024xll, Niemi:2024vzw, Wang:2024slx}, often in comparison with lattice studies~\cite{Gould:2021dzl, Ekstedt:2022zro, Gould:2023ovu, Gould:2023jbz, Gould:2024chm, Ekstedt:2024etx,Ramsey-Musolf:2024ykk}.

In section~\ref{sec:framework} we give an overview of the theoretical framework for dimensional reduction and bubble nucleation, which serves as a basis for the derivation of the catalog. In section~\ref{sec:catalog} we construct the catalog, with some derivations relegated to appendix~\ref{app:pc}. In section~\ref{sec:paramscan} we show the results of scanning a global likelihood over the Higgs sector SMEFT operators, revealing parts of the SMEFT parameter space in which a first-order phase transition is possible. In section~\ref{sec:pheno} we discuss various connections to phenomenology and gravitational wave detection, and we conclude with a discussion in section~\ref{sec:discussion}.

%================================================================================================================
\section{Theoretical framework}\label{sec:framework}
%================================================================================================================

\noindent{}A field trapped in a metastable minimum, separated from the global true minimum by a barrier, can escape to the true minimum either via quantum tunneling or via thermal escape. For early-universe phase transitions, we are interested in transitions at large temperatures and will hence focus on bubble nucleation via thermal escape. The dynamics of a scalar field undergoing such a transition with bubble nucleation take place at an energy scale much lower than the scale of thermal fluctuations. In more physical terms, the expected radius of the nucleated bubbles should be much larger than the length scale of thermal fluctuations, which is proportional to the inverse temperature. On the other hand, dynamics at energies higher than the thermal fluctuations is Boltzmann suppressed and can thus be ignored.

This separation of scales allows us to use the theoretical framework of Dimensional Reduction (DR), which yields a three-dimensional (3D) Euclidean field theory with only bosons. The first step in DR is to decompose the fields into their Matsubara modes due to their periodicity in the Euclidean time direction. Specifically, the fields are expanded as:
\begin{equation}
    \label{eq:boson_matsubara}
    \phi(\tau, \mathbf{x}) = \sum_{n=-\infty}^\infty e^{i \omega_n \tau} \phi_n(\mathbf{x}),
\end{equation}
where 
\begin{equation}
    \label{eq:boson_freq}
    \omega_n = 2 n \pi  T \, \text{ (bosons)}, \quad  \omega_n = (2 n + 1)\pi T \, \text{ (fermions)}, \quad n \in \mathbb{Z}.
\end{equation}
Each mode has a mass of \( m^2_n(T) = \omega_n^2 + m^2 \), where \( m \) is the mass at zero temperature. For very high temperatures, all but the bosonic \( n=0 \) modes are thus very massive. The second step is to integrate out the non-zero Matsubara modes,  obtaining an effective 3D theory of bosonic Matsubara zero-modes~\cite{Kajantie:1996mn,Farakos:1994kx}; fermions do not have zero-modes and are not present in the 3D theory. The zero-modes are interacting strongly at high temperatures and are hence the most important degrees of freedom. Notably, the 3D theory does not contain any temperature dependence, as the temperature dependence is encoded in relations between 3D and 4D parameters determined by matching to the full theory at the hard scale. In particular, this matching introduces Debye masses of the bosons, which contain a temperature dependence, while the potential does not explicitly depend on the temperature.

At this point, it is conceptually important to note the separation of scales involved in our framework. The scale of the thermal fluctuations, \( p \sim \pi T \), is referred to as the hard scale. It it the scale at which non-zero Matsubara modes are integrated out during DR. 
Integrating out the thermal fluctuations at the hard scale gives a perturbative field theory, which properly resums large logarithms~\cite{Gould:2021oba} (See~\cite{Gould:2021dzl} for a modern review of DR.) and captures the physics of the scalar field driving the phase transition. Well below this lies the scale where the phase transition occurs, characterized by the dynamics of the scalar field driving the transition, which is assumed to have a mass well below the hard scale. 

The assumption of a scalar field with a small mass compared to the temperature allows for the use of classical nucleation theory to describe the process of bubble nucleation~\cite{Gould:2021ccf}. This theoretical framework can be pushed to higher orders in perturbation theory without violating consistency conditions such as gauge invariance and realness of observables~\cite{Hirvonen:2021zej, Ekstedt:2022tqk, Lofgren:2021ogg, Lofgren:2023sep}.

As we will see, depending on the specific parameters of the 3D theory, there could be additional scales where some of the bosons in the 3D theory become relevant and can be integrated out (leading to yet another effective theory), leaving only the scalar field that drives the phase transition. Conceptually, this introduces intermediate scales between the hard scale and the phase transition scale, but the overall framework remains similar.

Recently, a hierarchy of scales has been properly introduced~\cite{Gould:2023ovu} to capture these nuances:
\begin{equation} \label{eq:scale-hierarchy}
    \underbrace{\vphantom{\frac{g^{\frac{1}{2}}}{\pi}}\pi T}_{\text{hard scale}}
    \gg
    \underbrace{\left(\frac{g}{\pi}\right)^{\frac{1}{2}} \pi T}_{\text{semisoft scale}}
    \gg
    \underbrace{\left(\frac{g}{\pi}\right)^{1} \pi T}_{\text{soft scale}}
    \gg
    \underbrace{\left(\frac{g}{\pi}\right)^{\frac{3}{2}} \pi T}_{\text{supersoft scale}}
    \gg
    \underbrace{\left(\frac{g}{\pi}\right)^2 \pi T}_{\text{ultrasoft scale}}
    \;.
\end{equation}
As we saw, the factor of \( \pi T \) comes from the scale of thermal fluctuations, and the factors of \( g/\pi \) come from loops within the 3D theory, where \( g \ll 1 \) is an appropriate power-counting parameter, in our case a gauge coupling in the original 4D theory.

For a first-order phase transition, the potential of the scalar field must contain two minima separated by a barrier. To understand how a barrier can arise in models with a \( \phi^6 \) term, we will primarily be interested in the potential of the Higgs field in the 3D effective theory left after DR,
\begin{align}
    V_3(\phi_3)&=\frac{1}{2}m_3^2 \phi_3^2+\frac{1}{4}\lambda_3 \phi_3^4+\frac{1}{6}\Cphithree\phi_3^6, \\
    [\phi_3]&=\frac{1}{2}, \quad [m_3^2]=2, \quad [\lambda_3]=1, \quad [\Cphithree]=0.
\end{align}

Note the dimensions of the different terms and components. Here \( \phi_3 \) is the scalar field in the 3D theory (the Matsubara zero-mode of the 4D scalar field \( \phi \), divided by \( \sqrt{T} \)),\footnote{Here we are glossing over the fact that the field \( \phi_3 \) is also properly normalized to get agreement between the screened masses of the 4D and 3D theories~\cite{Kajantie:1995dw}.} and \( m_3^2, \lambda_3 \), and \( \Cphithree \) are Wilson coefficients determined by matching to the full theory at high temperature. 

As all the temperature dependence is restricted to the matching relations, we can view the Wilson coefficients as functions of temperature. The phase portrait of the theory can be understood just in terms of these 3D Wilson coefficients: as the temperature changes, a particular path is traced in the phase portrait, depending on the values of the original couplings.

The simplest way for this theory to have a potential that contains two minima separated by a barrier is if the parameters satisfy \( m_3^2>0, \lambda_3<0, \Cphithree>0 \). Then a barrier is present at tree level within the high-temperature 3D EFT. But if such a tree-level barrier is not present, could one be generated by loop corrections?

To explore this question, it is useful to consider the hierarchy of scales introduced earlier. The dynamics at energies higher than the thermal fluctuations is Boltzmann suppressed and can thus be ignored. 

Assume now that the dynamics of the transitioning field take place at one of the scales below the hard scale, say the supersoft scale. If there are fields with dynamics at a higher scale, such as the soft scale, those fields can be integrated out to yield a nucleation EFT at the supersoft scale—possibly generating a barrier. The hierarchy of scales can protect the perturbative expansion, enabling us to study such radiative phase transitions perturbatively~\cite{Gould:2021ccf,Ekstedt:2022tqk}.

Thermal bubble nucleation can be understood as the process of a field undergoing thermal escape, where the scalar field is locally excited by thermal fluctuations and can escape over the barrier, passing through the field configuration of a critical bubble. 

Formulating this calculation as a process of thermal escape in a nucleation EFT is a consistent framework for calculating higher-order corrections. It has been shown to give gauge-invariant results in cases with radiative barriers (generated by loop corrections)~\cite{Hirvonen:2021zej, Lofgren:2021ogg}, and software has been developed to calculate the contributions of fluctuation determinants (1-loop contributions within the nucleation EFT)~\cite{Ekstedt:2023sqc}. 

The critical bubble configuration is found by extremizing the effective action of the nucleation EFT, which describes dynamics at the length scale defined by the size of the bubbles. This leads to a formula for the nucleation rate of the form

\begin{align}
    \Gamma_{\mathrm{nucl}}&=A_{\mathrm{dyn}}e^{-\bar{S}_{\mathrm{eff}}} \, ,\label{eq:nucleation-rate}\\
    \bar{S}_{\mathrm{eff}}&=\int_{x}\left[\frac{1}{2}(\partial \phi)^2+V_3(\phi)\right]+\cdots\label{eq:nucleation-action}
\end{align}
Here, the dynamical part of the rate, \( A_{\mathrm{dyn}} \), represents dissipative processes in the plasma and will contribute higher-order corrections to the rate. The bar in \( \bar{S}_{\mathrm{eff}} \) denotes that this effective action must be calculated with the negative eigenvalues of the transitioning field removed. This is to ensure that the saddle-point approximation is performed correctly: we should only integrate over the directions perpendicular to that of the critical bubble (which lies along the negative-eigenvalue direction). The critical bubble is then found by extremizing this action order by order in perturbation theory, with the leading order given by the action explicitly stated in \eqref{eq:nucleation-action}. The hierarchy of scales inherent in the construction of \( \bar{S}_{\mathrm{eff}} \) ensures that this is a meaningful procedure.

Let us illustrate these ideas with the example of the \( \text{SU}(2) \)+Higgs theory~\cite{Farakos:1994kx}, which we can take as a simplified version of the Standard Model electroweak sector. Our convention is that quantities in 3D carry the subscript 3. We then have
\begin{align}\label{eq:Lsu2}
    \mathcal{L}_{3,\text{scalar}}&=D_i \varphi_3^{\dagger} D_i \varphi_3 + m_3^2 \varphi_3^{\dagger} \varphi_3 + \lambda_3\left(\varphi_3^\dagger \varphi_3\right)^2,\\
    D_i \varphi_3 &= \left(\partial_i - i \frac{g_3}{2} \tau^a A_{3,i}^a \right) \varphi_3,\\
    \varphi_3&=\frac{1}{\sqrt{2}}\begin{pmatrix}
        \Phi_3^+\\
        \frac{1}{\sqrt{2}}\left( \phi_3 + i \Phi_3^0 \right)
    \end{pmatrix},\\
    [A_{3,i}^a]&=\frac{1}{2},\quad [g_3]=\frac{1}{2}.
\end{align}
Here, \( \tau^a \) are the Pauli matrices, \( A^a_{3,i} \) are the three spatial modes of the 3D gauge boson,\footnote{The fourth mode—the temporal scalar mode—has been integrated out to modify the Wilson coefficients of the theory. The separation of the temporal and spatial modes occurs because the heat bath breaks the Lorentz symmetry, which manifests as the compactified time dimension in dimensional reduction. The temporal mode can be treated as an independent scalar field with a thermal mass at the soft scale.} and \( \Phi_3^0, \Phi_3^\pm \) are Goldstone bosons. Note also that the 3D theory is Euclidean, hence the Latin index \( i \) and the plus signs in the last two terms. We will not be concerned with the individual components of \( \Phi_3 \), and can focus on the single field \( \phi_3 \) (treating it as a background field), to understand the phase structure,
\begin{align}
    V_3(\phi_3)&=\frac{1}{2}m_3^2\phi_3^2+\frac{1}{4}\lambda_3\phi_3^4,\\
    m_A^2&=\frac{1}{4}g_3^2\phi_3^2,\\
    n_A&=6.
\end{align}
The number of modes of \( A_{3,i}^a \) is \( n_A=2\times 3 \) because there are two polarization modes for each of the three SU(2) components of the gauge boson. This potential does not have a barrier, but one can be generated by integrating out the gauge boson.

What does this mean? As the gauge boson is massless unless the scalar field has a non-zero value, we need to integrate out the gauge boson in the presence of the scalar field's background field. This will generate non-polynomial terms in the effective potential. Such terms cannot arise in the usual way of constructing effective field theories, where one writes down polynomial terms and matches Wilson coefficients. But they can arise if one uses the technique of functional matching~\cite{Cohen:2020fcu} (see also~\cite{Hirvonen:2022jba})—also known as the background-field method~\cite{Abbott:1981ke}.

A familiar example of this idea is that of the Coleman-Weinberg potential~\cite{Coleman:1973jx}, where one integrates out a gauge field in the presence of a background field \( \phi \), yielding a potential that includes a logarithmic term \( \sim \log \phi \). When integrating out this gauge field with mass \( \sim g \phi \), care must be taken as this construction is only valid for a certain size of the background field \( \phi \). The effective action, and observables like the tunneling rate, are computed using a derivative expansion in powers of \( p^2 / (g^2 \phi^2) \). Hence it is necessary to ensure that the derivative expansion gives sensible results (see, e.g.,~\cite{Gould:2021ccf, Hirvonen:2021zej}), or the theory will not be perturbative.

To further demonstrate this idea, note that the 1-loop contribution to the effective potential of a bosonic mode within the 3D theory is given by~\cite{Kajantie:1993ag}
\begin{equation}
    f(m^2)\equiv\frac{1}{2}\int_{p}\log\left[p^2+m^2\right] =-\frac{1}{12\pi}\left(m^2\right)^{3/2}+\mathcal{O}(\epsilon),
\end{equation}
with the convention
\begin{equation}
    \int_{p}=\int\frac{\mathrm{d}^d p}{(2\pi)^d}, \qquad d=3-2\epsilon.
\end{equation}
The 1-loop contribution of the gauge boson,
\begin{align}
    n_A \times f(m_A^2)=6 \times \frac{-1}{12\pi}\left(\frac{1}{4}g_3^2\phi_3^2\right)^{\frac{3}{2}}=-\frac{1}{16\pi}g_3^3\phi_3^3,
\end{align}
would hence modify the leading-order potential to 
\begin{align}
    V_{\mathrm{LO}}(\phi_3)&\overset{?}{=}\frac{1}{2}m_3^2 \phi_3^2-\frac{1}{16\pi}g_3^3\phi_3^3+\frac{1}{4}\lambda_3 \phi_3^4+\frac{1}{6}\Cphithree\phi_3^6, \\
    [g_3]&=\frac{1}{2}.
\end{align}
The \( \phi_3^3 \) term gives rise to a barrier in the potential.\footnote{Writing the loop term in terms of \( \phi_3 \) makes the potential easier to work with, but it does hide the non-polynomial nature of the effective action. It also does not make gauge invariance clear. If we wanted to be more transparent, we would write the potential as a function of \( \Phi_3 \), with the \( \phi_3^3 \) term corresponding to \( (\Phi_3^\dagger \Phi_3)^{3/2} \).} The question mark above the equals sign signifies that this potential is only a valid perturbative description for certain sizes of the involved couplings and the scalar field \( \phi_3 \).

It is an exercise in power counting to determine what sizes of couplings allow for a radiative barrier, first performed in~\cite{Arnold:1992rz} and emphasized more recently in, e.g.,~\cite{Hirvonen:2021zej, Lofgren:2021ogg, Ekstedt:2020abj, Ekstedt:2022zro}. We review this exercise in the next section, where we catalog different barriers. Here we note that the result of the power counting, in the radiative barrier case described above, translated to the modern language of 3D EFTs, requires that the dynamics of the gauge field take place at the soft scale and the dynamics of the transitioning Higgs field at the supersoft scale~\cite{Gould:2023ovu}.

%================================================================================================================
\section{The Catalog}\label{sec:catalog} 
%================================================================================================================
With a theoretical framework set up to study first-order phase transitions within gauge theories with a $\phi^6$ term, we are now ready to explore the different possibilities. We will proceed by using power-counting methods to see what is implied by balancing the different terms. This is implemented by assuming a scaling with respect to the power-counting parameter $g$ for the different components of the potential:
\begin{align}
    m_3^2 &\sim g^{n_{m^2}} T^2,\label{eq:nm2}\\
    \lambda_3 &\sim g^{n_\lambda} T,\label{eq:nlambda}\\
    \phi_3 &\sim g^{n_\phi} \sqrt{T},\label{eq:nphi}\\    
    \Cphithree &\sim g^{n_C}.\label{eq:nc}
\end{align}
We will take the gauge coupling $g$ of our gauge theory to be our power-counting parameter, and we will in this section study the simpler SU(2)+Higgs+$\phi^6$ theory as preparation for the full SMEFT in the next section.
We do not include factors of $\pi$ here, as they are not important for deriving perturbative schemes. We will also assume that the corresponding 3D gauge coupling $g_3$ scales as $g_3 \sim g \sqrt{T}$, which follows from the leading order matching relations~\cite{Farakos:1994kx}. Then by balancing the different terms in the potential, we can look for solutions among the different $n$'s that allow the different terms to be of sufficient size. There are also extra constraints on these powers coming from perturbativity. See appendix~\ref{app:pc} for the details.

We can demonstrate the method on the simpler example of $\mathrm{SU}(2)$+Higgs introduced in the previous section. This model cannot have a barrier at tree-level, so let us consider the case of a radiative barrier:
\begin{equation}
    V(\phi_3)=\frac{1}{2}m_3^2 \phi_3^2-\frac{g_3^3}{16\pi}\phi_3^3+\frac{1}{4}\lambda_3 \phi_3^4.
\end{equation}
Balancing the terms here gives
\begin{align}
    m_3^2 \phi_3^2 &\sim g_3^3\phi_3^3 \sim \lambda_3 \phi_3^4,\nonumber\\
    \implies g^{n_{m^2}+2n_\phi} &\sim g^{3+3n_\phi} \sim g^{n_\lambda+4n_\phi},\nonumber\\
    \implies n_{m^2}+2n_\phi &= 3 + 3n_\phi = n_\lambda + 4n_\phi,\nonumber\\
    \implies n_{m^2}&=3+n_\phi, \quad n_\lambda = 3-n_\phi.
\end{align}
We see that the power-counting scheme allows for a family of solutions. We can constrain the values of $n_\phi$ by using the perturbativity conditions,
\begin{alignat}{2}
    V(\phi_3)&\gg g^6 T^3, \qquad&\text{Perturbativity},\\
    g_3^2 \phi_3^2 &\ll T^2, \qquad&\text{High-T expansion for gauge boson},\\ 
    m_3^2 &\ll T^2, \qquad&\text{High-T expansion for scalar}.
    %m_3^2 &\ll g_3^2 \phi_3^2,\qquad& \text{Hierarchy between gauge boson and scalar}.    
\end{alignat}
As we discuss in appendix~\ref{app:pc}, the first constraint is necessary to have a leading order dominant over the nonperturbative contributions of the ultrasoft scale. The two other constraints ensure that the high-T expansion applies to the included fields.

The perturbativity constraints then yield
\begin{align}
    3+3n_\phi < 6 &\implies n_\phi < 1,\\
    1 + n_\phi > 0 &\implies n_\phi > -1,\\
    3+n_\phi > 0 &\implies n_\phi > -3.
    %3+\nphi  >2+2\nphi &\implies \nphi < 1. 
\end{align}
Which resolves to
\begin{equation}
    -1 < n_\phi < 1.
\end{equation}
Any value within this range will lead to the same physical picture, and we will follow~\cite{Arnold:1992rz} and pick the geometric mid-point $n_\phi=0$. 

Our power-counting scheme hence is
\begin{align}
    m_3^2 &\sim g^3 T^2,\\
    \lambda_3 &\sim g^3 T,\\
    \phi_3 &\sim \sqrt{T}.
\end{align}
Note that a radiative barrier requires a relatively small quartic coupling in the 3D EFT, in contrast to the common assumption of $\lambda_3\sim g^2 T$.

This scheme encodes a particular hierarchy of scales within the 3D EFT. Namely, the transitioning scalar field is dynamic at the supersoft scale, while the spatial modes of the gauge boson are dynamic at the soft scale. Reinserting the factors of $\pi$ from \eqref{eq:scale-hierarchy},
\begin{equation}
    m_3^2 \sim \left(\frac{g}{\pi}\right)^3 \pi^2 T^2 \ll g_3^2 \phi_3^2 \sim \left(\frac{g}{\pi}\right)^2 \pi^2 T^2.
\end{equation}
As this radiative barrier arises through loops, this hierarchy is necessary for the perturbative expansion to be well-behaved.

This exercise in power counting was relatively straightforward. With more possible coefficients in the potential, the counting can become more complicated. In the next subsection, we catalog the result of balancing the possible terms in $\mathrm{SU}(2)$+Higgs$+\phi^6$. More details can be found in appendix~\ref{app:pc}.

\subsection{Overview of the catalog}\label{ssec:catOverview}

\begin{figure}
    \centering
    \includegraphics[width=\linewidth]{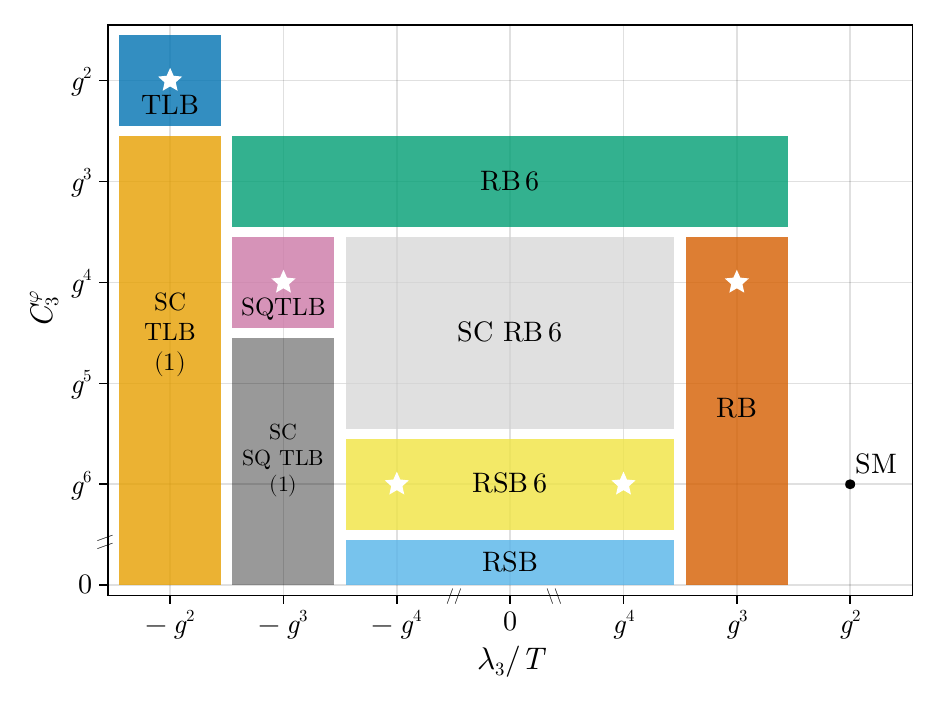}
    \caption{An overview of how the 3D parameter space can be sliced into different power-counting regions. Each region matches an entry in the catalog, in table~\ref{tab:catalog}. The white stars denote power-countings for which the SMEFT is estimated to predict the correct Higgs mass and vev, see subsection~\ref{ssec:param-mapping}. The black dot indicating the SM is based on lattice studies~\cite{Kajantie:1996mn} together with the DR matching relations. On the axes, note the "logarithmic" scale, and the hatch marks which emphasize that the axes have been cut. Finally, note that these regions are a projection which might lose some information, as the values of $m_3^2$ are not included, and that certain supercooled countings have been excluded as they overlap with other regions. As such, the figure should only be taken as a guiding sketch.     \label{fig:fopt-regions-3d}}
\end{figure}

We now extend the above example analysis to a gauge theory with a $\phi_3^6$ term, which yields a number of possible potentials with barriers. We list these in tables~\ref{tab:shorthands} and \ref{tab:catalog}, and illustrate this catalog its power counting relations in figure~\ref{fig:fopt-regions-3d}. The content of table~\ref{tab:catalog} is organized according to the hierarchy of scales needed for the particular potential to be a viable perturbative description. The different potentials are represented by the shorthand names shown in table~\ref{tab:shorthands}, and more details about each case is given in section~\ref{ssec:catDetails}.

\begin{table}
    \centering
        \begin{tabular}{ c|l|l }
        \hline 
        Shorthand & Meaning of acronym & Scale of Higgs dynamics \\
        \hline
        TLB & tree-level barrier & soft  \\
		\sqtlb & small quartic TLB & soft  \\
		RB & radiative barrier & supersoft  \\
		RSB & radiative symmetry breaking & soft \\
		\rbsix & RB with  dimension-six term $\phi_3^6$ & supersoft  \\
		\rsbsix & RSB with dimension-six term $\phi_3^6$ & soft  \\ 
		SC & supercooled variant & \\
        \hline 
        \end{tabular}
        \caption{Explanation of the shorthands in the catalog. The SC variants have different scale hierarchies than their regular variants, see table~\ref{tab:catalog}.}\label{tab:shorthands}
\end{table}

\begin{table}
    \centering
        \begin{tabular}{ c|c|c|c|c|r }
        \hline 
        Hierarchy & Shorthand & $\lambda_3 /T$ &$m_3^2 / T^2$ & $\phi_3/\sqrt{T}$ & $\Cphithree$\\ 
        \hline    
        \multirow{ 3}{*}{$m_3 \sim M \ll \pi T$} & TLB & $g^2$ & $g^2$ & $g^0$ & $g^2$ \\ 
        %\hline
        & SC TLB (1) & $g^2$ & $g^{2}$ & $g^{0}$ & $<g^2$ \\ 
        & SC TLB (2) & $g^2$ & $g^{3}$ & $g^{\frac{1}{2}}$ & $ g^2$ \\         
        \hline        
        \multirow{7}{*}{$m_3 \ll M \ll \pi T$} & \sqtlb{} & $g^3$ & $g^{2}$ & $g^{-\frac{1}{2}}$ & $ g^4$ \\
        & SC \sqtlb{} (1) & $g^3$ & $g^{2}$ & $g^{-\frac{1}{2}}$ & $<g^4$ \\          
        & SC \sqtlb{} (2) & $g^3$ & $g^{\frac{5}{2}}$ & $g^{-\frac{1}{4}}$ & $g^4$ \\          
        & RB & $g^3$ & $g^3$ & $g^0$ & $< g^3$ \\ 
        & \rbsix  & $g^3$ & $g^3$ & $g^0$ & $g^3$ \\         
        %\hline
        & SC RB & $g^3$ & $g^{\frac{7}{2}}$ & $g^{\frac{1}{2}}$ & $< g^{\frac{3}{2}}$ \\ 
        & SC RSB & $g^4$ & $g^{3}$ & $ g^{0}$ & $ < g^3$ \\                    
        \hline                          
        \multirow{ 2}{*}{$m_3 \ll M \sim \pi T$} & RSB & $g^4$ & $g^2$ & $ g^{-1}$ & $ < g^6$ \\  
        %\hline    
        & \rsbsix & $g^4$ & $g^2$ & $ g^{-1}$ & $ g^6$ \\      
        \hline 
        \end{tabular}
        \caption{A catalog of the different ways to generate a barrier in a gauge theory with a $\phi^6$ term, and the required scaling of the components of the potential. The Hierarchy column gives the corresponding scale hierarchy between the mass $m_3$ of the transitioning scalar field, the mass $M$ of the gauge boson, and the hard scale $\pi T$. Each case, written in shorthand here, is described in detail in subsection~\ref{ssec:catDetails}; SC means supercooled, and (1) or (2) refer to different power countings leading to the same effective potential, as described in Appendix~\ref{app:pc}.
        \label{tab:catalog}}
\end{table}

We note that many of the potentials require a value of $m_3^2$ which is small. In order for $m_3^2$ to be parametrically smaller than $g^2 T^2$, a cancellation needs to take place. From the leading order matching conditions in subsection~\ref{ssec:param-mapping}, we know that $m_3^2 =-m^2 + \# g^2 T^2$ where $m^2$ is the $\phi^2$ coefficient in the original theory, \# denotes a numeric coefficient, and $g^2$ represents the various couplings that contribute to this coefficient. 
For a cancellation to take place, we will need $m^2>0,m^2\sim g^2 T^2$. This is what happens for models with classical spontaneous symmetry breaking, and it implies that there exists a temperature $T_0$ for which the coefficient $m_3^2$ is zero: $m_3^2(T_0)=0$. This temperature would signal a second-order phase transition, and hence a candidate first-order phase transition needs to complete at a temperature above $T_0$. (The barrier disappears for temperatures below $T_0$, because there $m_3^2<0$.)

\subsubsection*{Supercooling}

Each potential also has one or more supercooled variants. Supercooling is the phenomenon where a system is cooled well below the critical temperature, yet it still stays in the metastable state. This can occur if the nucleation rate is not particularly large when the minima are close to degenerate. As time passes and the minima become more and more separated, a larger latent heat can be released in the transition.

The phenomenon of supercooling can be directly reflected in the power-counting schemes, by giving meaning to the size of the energy difference between the minima. When the difference becomes large enough, the true minimum can be neglected compared to the metastable minimum, and the potential can be modeled as a potential that is unbounded from below. The supercooled potentials we consider all have this feature, and either feature smaller $m_3^2$---and thus larger bubbles---or a smaller $\phi_3^6$ coefficient
than produced by the transitions of the corresponding potentials from which they supercooled. This reflects that the balance between the barrier and the metastable minimum occurs for smaller field values during supercooling.

Somewhat counter-intuitively, potentials unbounded from below, but with a barrier, do not lead to an infinite decay rate. Though the energy difference between the vacua is formally infinite, at the moment of nucleation the region inside of the critical bubble is not of the true vacuum but of the value at the escape point (the field value at the origin found by solving the bounce equation). Hence the energy difference is finite and so is the nucleation rate. This is similar to the unbounded from below, and scale-free, potentials described in~\cite{Ekstedt:2023sqc}.\footnote{Similar considerations regarding supercooled nucleation in terms of power counting are discussed in~\cite{Gould:2021ccf}.}

To be more concrete, we can illustrate power-counted supercooling by continuing with our example of a radiative barrier (RB) in $\mathrm{SU}(2)$+Higgs. For large temperatures the field is trapped in the symmetric minimum at the origin,
\begin{equation}
    V(\phi_3)=\frac{1}{2}m_3^2 \phi_3^2 + \frac{1}{4}\lambda_3 \phi_3^4, \qquad m_3^2 \sim g^2 T^2,\quad m_3^2>0, ~\lambda_3 > 0.
\end{equation}
But as the temperature lowers and approaches $T_0$, $m_3^2=-m^2 + \# g^2 T^2$ becomes smaller and a barrier can form. A new minimum comes into existence and a phase transition becomes possible,
\begin{equation}
    V(\phi_3)=\frac{1}{2}m_3^2 \phi_3^2-\frac{g_3^3}{16\pi}\phi^3+\frac{1}{4}\lambda_3\phi_3^4, \qquad m_3^2 \sim g^3 T^2.
\end{equation}
As the temperature lowers even further, if the field is still trapped in the metastable symmetric minimum (e.g. due to a low nucleation rate), the system supercools: $m_3^2$ becomes even smaller and the true minimum becomes negligible from the point of view of the scalar field at the symmetric minimum,
\begin{equation}
    V(\phi_3)=\frac{1}{2}m_3^2 \phi_3^2-\frac{g_3^3}{16\pi}\phi_3^3, \qquad m_3^2 \sim g^{3\frac{1}{2}} T^2.
\end{equation}
As described above, though the potential is unbounded from below, the nucleation rate is still finite. The dynamics of the bubble are set by the size of the barrier in comparison to the mass of the field. If the temperature lowers yet more, without nucleation occurring, $m_3^2$ becomes negative and the barrier is eliminated,
\begin{equation}
    V(\phi_3)=-\frac{1}{2}|m_3^2| \phi_3^2+\frac{1}{4}\lambda_3\phi_3^4, \qquad |m_3^2| \sim g^{2} T^2.
\end{equation}
Then a smooth phase transition takes place instead.

\subsection{Details of the catalog}\label{ssec:catDetails}

In this subsection, we list the different potentials in the catalog in table~\ref{tab:catalog}, and note some restrictions on the parameters for a barrier to be able to form. We only include those terms in the potential that are of sufficient size to contribute to the formation of the barrier.

\subsubsection*{Tree-Level Barrier (TLB and \sqtlb)}
With a negative $\lambda_3$, it is possible to get a barrier just from the tree-level terms,
\begin{align}\label{eq:TLBpot}
    V_3(\phi_3)&=\frac{m_3^2}{2} \phi_3^2-\frac{|\lambda_3|}{4} \phi_3^4+\frac{\Cphithree}{8}\phi_3^6,\\
    m_3^2 &> 0, \quad\lambda_3 < 0, \quad\Cphithree > 0.\nonumber
\end{align}
To find the critical temperature $T_c$ at leading order, we would need to first find the critical mass of this 3D potential. The critical mass $(m_3^2)_c\equiv (m_3^2)|_{T_c}$ is the value of $m_3^2$ where the broken minimum is degenerate with the symmetric minimum. Making the broken minimum degenerate with the symmetric minimum, and using that $dV/d\phi = 0$ at the minima gives 
\begin{equation}\label{eq:criticalmassTLB}
   \left(m_3^2\right)_{c}=\left.\frac{\lambda_3^2}{4\Cphithree}\right|_{T_c}.
\end{equation}
Then the leading order value of $T_c$ can be found by solving this equation after inserting the temperature dependence of these Wilson coefficients from the matching relations of subsection~\ref{ssec:param-mapping}. 

It is actually possible to realize this same potential with two different power-countings. The first is dubbed TLB (short for tree-level barrier), and features a quartic coupling of size $\lambda_3\sim g^2 T$. This size of $\lambda_3$ is the same as is used in ordinary loop expansions.

Perhaps more surprising is the existence of a second counting, \sqtlb{} (Small Quartic TLB), which features $\lambda_3\sim g^3 T$ among its other parameter scalings. That size of $\lambda_3$ often implies that a radiative barrier can form~\cite{Arnold:1992rz}. In this case, there is a hierarchy between the scalar and the gauge boson, with the scalar at the soft scale and the gauge boson at the semisoft scale. But still there is no radiative barrier, because the $\phi_3^3$ term is subleading. We can then draw the conclusion that a hierarchy between the scalar and the gauge boson is necessary, but not sufficient, for a radiative barrier to arise.

\subsubsection*{Supercooled (SC) TLB}
If the nucleation rate for the TLB potential is small, the field might stay in the metastable minimum until supercooling happens. Then the potential will take the form
\begin{align}\label{eq:SCTLBpot}
    V_3(\phi_3)&=\frac{m_3^2}{2} \phi_3^2-\frac{|\lambda_3|}{4} \phi_3^4,\\
    m_3^2 &> 0, \quad\lambda_3 < 0.\nonumber
\end{align} 
By construction, this potential is unbounded from below and only features the metastable minimum. It can only describe the physical system after it has cooled beyond the critical temperature, and can therefore be used to calculate either the nucleation temperature or the percolation temperature (which are both lower than $T_c$). However, it is not possible to find $T_c$ from this potential, although it can be found by studying the original TLB potential. 

\subsubsection*{Radiative barrier (RB)}
As described above, if the scalar field is lighter than the gauge boson, then the gauge boson can be integrated out. This results in a radiative barrier, with potential
\begin{align}\label{eq:RBpot}
    V_3(\phi_3)&=\frac{m_3^2}{2} \phi_3^2-\frac{g_3^3}{16\pi}\phi_3^3+\frac{\lambda_3}{4} \phi_3^4,\\
    m_3^2 &> 0, \quad\lambda_3 > 0.\nonumber
\end{align}
The critical mass is
\begin{equation}
    \left.m_3^2\right|_{T_c}=\left.\frac{g_3^6}{128 \pi^2 \lambda_3}\right|_{T_c}.
\end{equation}
\subsubsection*{Radiative barrier with $\phi^6$ term (\rbsix)}
If a radiative barrier is generated, and the Wilson coefficient $\Cphithree$ is large enough, then they can both play a role in the formation of a barrier, as it would allow for a negative $\lambda_3$:
\begin{align}\label{eq:RB6pot}
    V_3(\phi_3)&=\frac{m_3^2}{2} \phi_3^2-\frac{g_3^3}{16\pi}\phi_3^3+\frac{\lambda_3}{4} \phi_3^4+\frac{\Cphithree}{8}\phi_3^6,\\
    m_3^2 &> 0, \quad\Cphithree > 0.\nonumber
\end{align}
The critical mass can be found from
\begin{align}
    \left(m_3^2\right)_{c}&=-\frac{3}{4}\Cphithree \phi_{3,c}^4+\frac{3 g_3^3 \phi_{3,c}}{16 \pi }-\lambda_3\phi_{3,c}^2,\label{eq:critMassRB6}\\
    \phi_{3,c}\left(\Cphithree \phi_{3,c}^2+\lambda_3\right)&=\frac{g_3^3}{8\pi},\label{eq:critPhiRB6}
\end{align}
where $\phi_{3,c}$ is the location of the broken minimum. It is possible to solve \eqref{eq:critPhiRB6} analytically, but we opt to not do so here for brevity.
\subsubsection*{Radiative symmetry breaking (RSB)}
Radiative symmetry breaking (RSB), first realized by the Coleman-Weinberg (CW) mechanism~\cite{Coleman:1973jx}, occurs when there is no symmetry breaking present in the classical potential, but instead the symmetry is broken when loop corrections are included. At zero temperature, this requires the scalar coupling to be small compared to the gauge-boson coupling, $\lambda \sim g^4$.

There are obstacles to studying RSB at finite temperature. First and foremost is the problem that the gauge bosons are too heavy for the high-T expansion to apply, posing challenges to the framework of DR.

However, as explained in~\cite{Lofgren:2023sep}, it is still possible to use a 3D EFT to describe the bubble nucleation---as long as the scalar is still light. Constructing the 3D EFT of the nucleating field hence requires integrating out the gauge boson at the same time as the thermal fluctuations are integrated out. This means we need to include the full 1-loop thermal functions in our LO potential. (See~\cite{Kierkla:2023von} for an alternate approach to studying RSB using DR.)
 
This results in an effective potential of the form
\begin{align}\label{eq:RSBpot}
    V_3(\phi_3)&=\frac{m_3^2}{2} \phi_3^2+\frac{\lambda_3}{4} \phi_3^4+\frac{9}{T}J_A(g_3 \phi_3)+9\frac{T^3}{2\pi^2}J_{A}^{T\neq 0}\left(\frac{g_3^2 \phi_3^2}{T^2}\right),\\
    m_3^2 &> 0, \quad\lambda_3 > 0.\nonumber
\end{align}
See \eqref{eq:CWIntegralFunc} and \eqref{eq:finTIntegralFunc} for definitions of the loop functions. The factor 9 comes from the physical modes of the gauge boson, three spacetime components for each of the three gauge components. This factor 9 is different from the radiative barrier case (with a factor of 6) because the spatial and temporal modes are only separated when the gauge boson is dynamic at some scale below the hard scale. Because the gauge boson mass is here as large as the hard scale, we should use the full 3 spacetime components.

The critical mass is not likely to be found analytically, and ought to be determined numerically.

\subsubsection*{Radiative symmetry breaking with $\phi^6$ term (\rsbsix)}
The RSB potential discussed above can be modified if the Wilson coefficient $\Cphithree$ is large enough. Then the potential takes the form
\begin{align}\label{eq:RSB6pot}
    V_3(\phi_3)&=\frac{m_3^2}{2} \phi_3^2+\frac{\lambda_3}{4} \phi_3^4+\frac{9}{T}J_A(g_3 \phi_3)+9\frac{T^3}{2\pi^2}J_{A}^{T\neq 0}\left(\frac{g_3^2 \phi_3^2}{T^2}\right)+\frac{\Cphithree}{8}\phi_3^6,\\
    m_3^2 &> 0, \quad\Cphithree > 0.\nonumber
\end{align}
Again, the critical mass should be determined numerically.

\subsubsection*{SC RB and SC RSB}
If the nucleation rate for the RB (or RSB) potential is small, the field might stay in the metastable minimum until supercooling happens. Then the potential will take the form
\begin{align}\label{eq:SCRBpot}
    V_3(\phi_3)&=\frac{m_3^2}{2} \phi_3^2-\frac{g_3^3}{16\pi}\phi_3^3,\\
    m_3^2 &> 0.\nonumber
\end{align}
In this case the high-T expansion does apply to the gauge boson, even if this is a supercooled RSB potential. This is because the supercooled potential is valid for smaller $\phi_3$ values than the non-supercooled potential.

This potential can only describe the physical system after it has cooled beyond the critical temperature. Hence, it is not possible to find $T_c$ from this potential.

%================================================================================================================
\section{Mapping the parameter space of the SMEFT}\label{sec:paramscan}
%================================================================================================================

In this section, we turn to the application of the catalog in table~\ref{tab:catalog} to the study of the Higgs sector of the SMEFT. We want to understand which parts of the parameter space can support a first-order phase transition, and whether those regions are compatible with experimental data. 

Note that we now return to the 4D theory---it is important to note that we constructed the catalog in table~\ref{tab:catalog} purely from the point of view of the 3D EFT that describes the scalar field. It is meant to primarily be an exploration of the possible kinds of barriers in gauge theories with a $\phi^6$ term. This means that we cannot guarantee that each counting can actually be realized by matching to the parameters of the corresponding 4D QFT, though we will explore this question in this subsection. We will now include the full SMEFT electroweak sector with the SU(2)$\times$U(1) gauge group and corresponding gauge couplings $g$ and $g'$, and in addition the Yukawa coupling $y_t$ of the top quark. These interactions are included in the parameter scans that will follow, and will appear in the matching conditions of the 3D and 4D theories.

\subsection{The setup}
We will use the same setup as in~\cite{Camargo-Molina:2021zgz}, namely we will use the Warsaw basis~\cite{Grzadkowski:2010es} in the notation of~\cite{Dedes:2017zog}, but we only consider the three dimension-six operators that are built exclusively from the Higgs doublet $\varphi$. These operators are
\begin{align}
    \Qphi &= (\varphi^\dagger \varphi)^3,\\
    \Qphibox &= (\varphi^\dagger \varphi)\Box(\varphi^\dagger \varphi),\\
    \QphiD &= (\varphi^\dagger D_\mu \varphi)^*(\varphi^\dagger D^\mu \varphi).
\end{align} 
The  Lagrangian is then given by 
\begin{equation}
    \mathcal{L}\supset(D_\mu \varphi)^\dagger(D^\mu \varphi)+m^2(\varphi^\dagger \varphi)-\frac{\lambda}{2}(\varphi^\dagger \varphi)^2+\Cphi \Qphi + \Cphibox \Qphibox +  \CphiD \QphiD.
\end{equation}
We expand the Higgs doublet in terms of components as
\begin{equation}
    \varphi = \begin{pmatrix}
        \Phi^+\\
        \frac{1}{\sqrt{2}}\left(\phi+H+i\Phi^0\right)
    \end{pmatrix},
\end{equation}
where $\phi$ is a generic background field. Because we are mainly interested in symmetry breaking and phase transitions, we will focus on $\phi$ and not the actual degrees of freedom $H, \Phi^0,$ and $\Phi^+$. In terms of this background field, the classical potential is
\begin{equation}\label{eq:classical-potentialSMEFT}
    V_0(\phi)=-\frac{m^2}{2}\phi^2+\frac{\lambda}{8}\phi^4-\frac{\Cphi}{8}\phi^6.
\end{equation}
To keep this potential bounded from below we will consider $\Cphi<0$, but $m^2$ and $\lambda$ can have any sign as long as a broken minimum exists.

The three non-zero Wilson-coefficients will modify the equations for the Higgs mass and the Higgs vacuum expectation value. The potential term $\Cphi \Qphi$ will contribute directly to both the mass and the vacuum expectation value. The derivative terms $\QphiD$ and $\Qphibox$ will modify the kinetic terms of the scalar fields, requiring field redefinitions of the scalar fields to assure they are canonically normalized. This introduces a dependence on $\CphiD$ and $\Cphibox$ into the Higgs mass. 

We are also interested in cases where $\lambda$ is very small in absolute value, $|\lambda|\sim g^4$. This is the realm of radiative symmetry breaking~\cite{Coleman:1973jx}, where quantum corrections can play a role in spontaneous symmetry breaking. We will consider the gauge bosons and the top quark as coupling strongly enough to the Higgs field to potentially give appreciable radiative corrections.

From \eqref{eq:classical-potentialSMEFT} and the tree-level mass in~\eqref{eq:treelevelHiggsmass} below, we can estimate the sizes of the involved couplings,
\begin{align}
    m^2&\sim \lambda v^2,\label{eq:pc-m2}\\
    \Cphi&\sim \frac{\lambda}{v^2},\label{eq:pc-Cphi}\\
    \CphiD&\sim \frac{1}{v^2},\label{eq:pc-CphiD}\\
    \Cphibox&\sim \frac{1}{v^2}.\label{eq:pc-Cphibox}
\end{align}
These estimates hint at the expected sizes of the coefficients for them to make relevant contributions to the Higgs mass and vacuum expectation values. In our previous work we found that the experimental constraints on $\CphiD$ and $\Cphibox$ actually require them to be much smaller~\cite{Camargo-Molina:2021zgz}. The coefficient $\Cphibox$ is of the same order as $\Cphi$, or smaller. The coefficient $\CphiD$ must be even smaller from constraints set by electroweak precision tests, primarily because it breaks custodial symmetry. 

Our approach in this paper will be to still include $\CphiD$ and $\Cphibox$ when calculating the Higgs mass and the vacuum expectation value, expecting their contributions to widen the viable parameter space somewhat but not dramatically. When estimating phase transition observables we will also assume that these derivative operators are small enough that they first contribute at next-to-leading order, which would not have been the case if $\CphiD\sim\Cphibox\sim1/v^2$.

We do not include dimension-8 operators in our analysis. Consider adding a term $-C^8 \phi^8$ to the potential \eqref{eq:classical-potentialSMEFT}. Using Eqs.~(\ref{eq:pc-m2}--\ref{eq:pc-Cphi}) it is then straightforward to see that if the Wilson coefficient $C^8$ takes its natural, large, value $\Lambda^{-4}$, it will be suppressed by one factor of $\lambda$ compared to the other terms in the potential. It will therefore not play a role in the formation of the barrier or in determining the value of the minimum. 

\subsection{Viable parameter space}\label{ssec:inputs}

The model has five free parameters, $\lambda$, $m^2$, $\Cphi, \CphiD$, and $\Cphibox$. As we will explain in this section, $\lambda$ and $m^2$ are fixed by requiring that the correct vacuum expectation value $v$ and Higgs mass $m_h^2$ are obtained. We therefore need to scan over three free parameters. 

In order to efficiently explore the parameter space of $\Cphi$, $\CphiD$, and $\Cphibox$, we use a genetic algorithm (GA) tailored for high-dimensional optimization problems with complex constraints. Our primary goal is not to perform a comprehensive statistical inference over the full SMEFT parameter space, but rather to identify viable parameter points that are consistent with current experimental constraints. Genetic algorithms are particularly well-suited for this task, being very good at rapidly finding good solutions in complex landscapes without requiring gradient information or extensive sampling of the entire parameter space.

We use the Python package \texttt{lightweight-genetic-algorithm}~\cite{Wessen:2024pzm} (written by one of us), which uses a diversity-enhanced selection mechanism in the GA. This allows a diverse population of solutions and prevents the algorithm from becoming trapped in local optima, thus ensuring a broad variety of viable parameter points.

Our scan ran over 10 generations with a population size of 5000 individuals. Each individual represents a set of values for the Wilson coefficients $\Cphi$, $\CphiD$, and $\Cphibox$, varied within the following ranges (in units of $\GeV^{-2}$):
\begin{align}
        \Cphi &\in [-8 \times 10^{-6},\, 0], \\
        \CphiD &\in [-5 \times 10^{-7},\, 5 \times 10^{-7}], \\
        \Cphibox &\in [-5 \times 10^{-6},\, 5 \times 10^{-6}].
\end{align}
These ranges were selected to include possible regions where the coefficients can have significant effects on Higgs physics without violating perturbativity or current experimental bounds.
    
In the GA, we employ the ``between'' crossover method for numerical genes~\cite{Wessen:2024pzm}, where offspring genes are randomly chosen between the corresponding genes of the parents. Mutations are introduced using a ``random'' mutation mode with a rate of 1 \% per gene per individual, wherein mutated genes are assigned new random values within their allowed ranges.
    
The fitness function of the GA, which drives the evolution of the population, is the global likelihood for each parameter point. We calculate the likelihood using the public code \texttt{smelli}~\cite{smelli}, which in turn uses the public codes \texttt{flavio}~\cite{flavio} to calculate observables and \texttt{wilson}~\cite{wilson} for renormalization group running, while \texttt{smelli} aggregates experimental data from electroweak precision tests, Higgs measurements, and other relevant observables. To accurately account for the effects of the dimension-six operators on the Higgs mass, we override the default functions in \texttt{wilson} by implementing custom matching conditions (fixing $\lambda$ and $m^2$), which include one-loop contributions from $\Cphi$, $\CphiD$, and $\Cphibox$.

We calculate these one-loop contributions to the Higgs mass with the help of a sequence of public codes. We have written a model file for \texttt{FeynRules}~\cite{FeynRules}, and have then used \texttt{FeynHelpers}~\cite{FeynHelpers}, which combines \texttt{FeynArts}~\cite{FeynArts}, \texttt{FeynCalc}~\cite{FeynCalc} and \texttt{PackageX}~\cite{Patel_2017}, to calculate the one-loop correction to the Higgs mass in the $\overline{\text{MS}}$ scheme, working in the Feynman–'t Hooft gauge. We have explicitly verified that the final expression is gauge invariant and matches the SM expression when the Wilson coefficients are set to zero.

The tree-level part of the mass in the presence of a background field $\phi$, the ``field-dependent mass,'' can be found from the potential in \eqref{eq:classical-potentialSMEFT}, together with a field renormalization $Z_h^{2}$ introduced to give canonically normalized kinetic terms (as described in~\cite{Dedes:2017zog}),
\begin{align}
    Z_h^{2}&=1+2C^{\text{kin}}\phi^2,\\
    C^{\text{kin}}&=\frac{\CphiD}{4}-\Cphibox.
\end{align}
The minimum of the tree-level potential $v$ found by minimizing \eqref{eq:classical-potentialSMEFT} is 
\begin{align}
v^2=\frac{2 m^2}{\lambda}+\frac{6 m^4}{\lambda^3}\Cphi
\end{align}
The tree-level mass evaluated at $\phi=v$ is then
\begin{align}\label{eq:treelevelHiggsmass}
    m_h^2=Z_h^{-2}\partial^2_\phi V_0(\phi=v)=-m^2+\frac{3 \lambda  v^2}{2}-2C^{\text{kin}}v^2\left(\frac{3 \lambda  v^2}{2}-m^2\right)-\frac{15}{4} \Cphi v^4,
\end{align}
where higher-order terms in the EFT expansion have been discarded.

In our one-loop calculation of the Higgs mass, we have performed a strict perturbative expansion by linearizing the one-loop corrections with respect to $\lambda$ and the dimension-six Wilson coefficients. That is, we calculate the pole mass perturbatively,
\begin{align}
    M_0^2&=m_h^2,  \\
    M^2&=M_0^2+\Pi(p^2=M^2)=M_0^2+\hbar\Pi_1(p^2=M_0^2)+\Ordo{\hbar^2},
\end{align}
where $m_h^2$ is given by \eqref{eq:treelevelHiggsmass}, $\hbar$ is a power-counting parameter, and $\Pi_1(p^2=M_0^2)$ is the 1-loop contribution we calculated using \texttt{FeynHelpers}. This procedure respects the assumed power counting underlying \eqref{eq:classical-potentialSMEFT}, and as such avoids spurious IR divergences.

In most input schemes used in the SMEFT, $\lambda$ is fixed in such a way that the SM relationship with the Higgs mass is preserved, order by order, but usually at tree-level \cite{Brivio_2021}. That means that what is called $\lambda$  by construction has the same value as in the Standard Model, independently of the values of Wilson coefficients. Here we have taken a slightly different approach and take the Higgs mass to be a predicted observable. This is because in our analysis it is important to understand the potential using the values of the Wilson coefficients and $\lambda$ and how they affect different power-counting schemes.

In other words, we mostly use the ($M_Z$, $\alpha$, $G_F$) input scheme, with one technical exception: We find $\lambda$ to ensure that the calculated Higgs mass matches the observed value of $125.09\,\text{GeV}$ within a tolerance of $1\,\text{GeV}$. With this approach then $\lambda$ takes values that are different from the SM in order to match the Higgs mass and be consistent with the input value of $v$.

This procedure adjusts $\lambda$ for each set of Wilson coefficients, ensuring accurate predictions for the Higgs mass at the one-loop level while accounting for the effects of the dimension-six operators. Besides the Higgs mass, the input values for the SM match \texttt{smelli}.

This way we relate input parameters to observables through matching conditions, consistent with standard SMEFT practices, so that our exploration of the parameter space using the GA yields viable points in the Wilson coefficient parameter space that respect experimental constraints and reflect the impact of new physics in the Higgs sector. 
    
\begin{figure}
    \includegraphics[width=\linewidth]{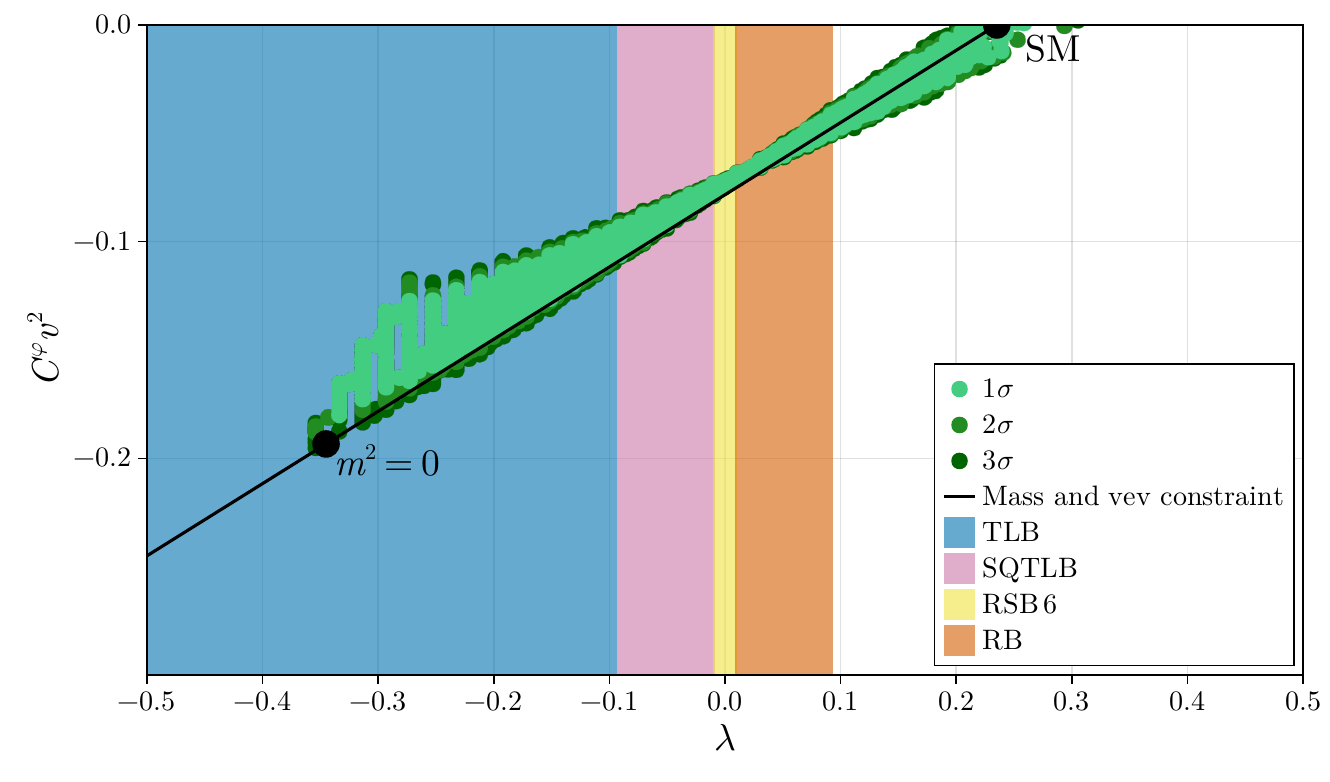}
    \caption{The points found by our scan, with better likelihood points on top. The points are colored according to whether they fall within the $1\sigma$, $2\sigma$, or $3\sigma$ confidence regions compared to the SM null hypothesis. 
    The black line is the simplified prediction for which parameters get the correct Higgs mass and vev, based on the  analysis in \eqref{eq:CphiPrediction}. The shaded regions reflect which power countings from table~\ref{tab:catalog} will be relevant for studying the phase transition in those regions, according to the analysis in sections~\ref{ssec:param-mapping} and \ref{ssec:first-order}.
    The boundaries demarcating these regions are leading-order estimates based on the analysis in Sec.~\ref{ssec:first-order}; they are expected to shift and theoretical uncertainties upon the inclusion of higher-order corrections.}  \label{fig:res-cphi-lambda}
\end{figure}

Given the computational intensity of evaluating the fitness function for a large population over multiple generations, we ran the GA on a high-performance computing instance with 89 cores provided by the Google Cloud Platform (GCP). The entire scan was completed in approximately one week. The results can be seen in figure~\ref{fig:res-cphi-lambda}.

From the point of view of the catalog in section~\ref{sec:catalog}, we will be interested in different sizes of the couplings $\lambda$ and $\Cphi$, as these are the main determinants of the corresponding couplings $\lambda_3$ and $\Cphithree$ in the 3D EFT. 

\subsection{Mapping 4D to 3D}\label{ssec:param-mapping}

Our aim is to understand which parts of the SMEFT parameter space can support a first-order phase transition. In the theoretical framework of DR, we can understand the behavior of our model as the temperature is changing via the temperature-dependence of the Wilson coefficients of the 3D EFT. As the temperature is decreasing, as is the case in the early universe, any particular parameter point of the SMEFT will pick out a path in the space of 3D Wilson coefficients.

From the catalog in table~\ref{tab:catalog} we also know that the form of the potential is determined by the relative sizes of the 3D Wilson coefficients. To determine whether a particular parameter point of the SMEFT has a first-order phase transition, we are then faced with the question of whether the corresponding path in the space of 3D Wilson coefficients enters a region where one of the power countings of the catalog can apply.

To attempt to answer this question, we can get an appreciation for the shape of the paths from the leading order matching conditions,
\begin{align}
    \phi_3 &= \frac{\phi}{\sqrt{T}}+\mathellipsis,\label{eq:phi3-matching-LO}\\
    m_3^2&=-m^2+\left(3\lambda+\frac{9}{4}g^2+\frac{3}{4}g'^2+3y_t^2\right)\frac{T^2}{12}-\frac{1}{4}\Cphi T^4+\mathellipsis,\label{eq:m32-matching-LO}\\
    g_3&=g \sqrt{T} +\mathellipsis,\label{eq:g3-matching-LO}\\
    g'_3&=g'\sqrt{T}+\mathellipsis,\label{eq:gp3-matching-LO}\\
    \lambda_3&=\frac{\lambda}{2} T - \Cphi T^3+\mathellipsis,\label{eq:lambda3-matching-LO}\\
    \Cphithree&=-\Cphi T^2+\mathellipsis,\label{eq:Cphi3-matching-LO}\\
    \CphiDthree&=\CphiD T^2+\mathellipsis.\label{eq:Cphi3D-matching-LO}\\
    \Cphiboxthree&=\Cphibox T^2+\mathellipsis,\label{eq:Cphi3box-matching-LO}    
\end{align}

Except for the final two relations in~\cref{eq:Cphi3D-matching-LO,eq:Cphi3box-matching-LO}, which were derived by hand, these relations were derived using the Mathematica code \texttt{DRalgo}~\cite{Ekstedt:2022bff}, and cross-checked with those in~\cite{Croon:2020cgk}. These matching relations can be extended beyond leading order, and in fact such exactness is necessary when performing high precision calculations. As we are here only interested in deriving power counting relations, we are content with not going beyond the leading order expressions. Any higher order corrections, being suppressed relative to the leading order, will not affect the power counting rules.

The formula for $m_3^2$ is not the same as the one we used for $m^2_{\mathrm{eff}}$ in~\cite{Camargo-Molina:2021zgz}: we here include the $\Cphi$ contribution, and neglect the $\CphiD$ and $\Cphibox$ contributions---a reflection of the power counting discussion after \cref{eq:pc-m2,eq:pc-Cphi,eq:pc-CphiD,eq:pc-Cphibox}. Inspecting the relations in~\cref{eq:phi3-matching-LO,eq:m32-matching-LO,eq:g3-matching-LO,eq:gp3-matching-LO,eq:lambda3-matching-LO,eq:Cphi3-matching-LO,eq:Cphi3D-matching-LO,eq:Cphi3box-matching-LO} tells us a few things. First, $g_3$ and $g'_3$ are proportional to their 4D counterparts. This allows us to estimate their sizes just from the sizes of the corresponding 4D quantities: $g_3/\sqrt{T}\sim g$.

The coefficients $m_3^2$ and $\lambda_3$ are somewhat more difficult, because it is possible for cancellations to take place that disconnect the sizes of the 3D parameters from the 4D parameters. We have already seen an example of this in section~\ref{ssec:catOverview} where we noted that $m_3^2$ can be smaller than $g^2 T^2$ due to cancellations between $-m^2$ and the high-temperature corrections $\sim g^2 T^2$. In principle the same could happen with $\lambda_3$, a possibility we explore in appendix~\ref{app:estimating-coeffs}. It turns out this possibility is not perturbatively viable, and would require fine-tuning of the parameters. We will hence ignore this scenario and simply estimate $\lambda_3/T \sim \lambda$.

Similarly, $\Cphithree$ is dimensionless but still depends on the temperature $T$. To estimate how the size of $\Cphithree$ depends on the zero-temperature coefficient $\Cphi$, we will hence need to estimate the temperature for which the phase transition takes place.

To estimate $T_c$ we will need to consider two separate cases. The first is with moderately sized $m^2$. In this case the phase transition will take place close to the cancellation temperature $T_0$, where $m_3^2(T_0)=0$. This allows us to use the estimation $m^2\sim g^2 T^2$, which together with the expected sizes of $\Cphi$ and $m^2$, gives
\begin{equation}
    \begin{rcases}
        m^2&\sim \lambda v^2\sim g^2 T^2\\
        \Cphi&\sim \lambda/v^2\\
    \end{rcases}
    \implies \Cphithree\sim \Cphi T^2 \sim \frac{\lambda^2}{g^2}.
\end{equation}
We use this estimation of $\Cphithree$ to mark regions in figure~\ref{fig:fopt-regions-3d} where it is expected to give the correct Higgs mass (as contributions from $\Cphi$ are large enough to contribute appreciably there).

On the other hand, some items in the catalog do not require the near cancellation of $m_3^2$. These cases could still occur if $m^2 \ll g^2 T^2$. Hence, we can also consider a Coleman-Weinberg-type (CW-type) scenario where $m^2=0$~\cite{Coleman:1973jx}. This scenario is interesting because the scale of the critical temperature is not set by balancing the terms in $m_3^2$. Indeed, the only scaleful parameter is $\Cphi \sim \lambda/v^2$, and the dominant temperature dependence comes from $\Cphithree$. To find the critical temperature we would solve \eqref{eq:criticalmassTLB}; using the estimate $\Cphi\sim\lambda/v$, gives
\begin{equation}
    T_{\mathrm{CW}}\sim v.
\end{equation}
Which allows us to estimate that 
\begin{equation}
    \left.\Cphithree\right|_{\mathrm{CW}} \sim \Cphi v^2 \sim \lambda.
\end{equation}
Adding to this analysis is the fact that if $m^2$ is too small to contribute to $m_3^2$, then $m_3^2\sim g^2 T^2$. Hence, the CW case must match to a 3D power-counting of the form $\nmSq=2, \nlambda = \nC$. Perusing the catalog in table~\ref{tab:catalog} suggests that the only option is TLB, a tree-level barrier. Furthermore, we can use \eqref{eq:CphiPrediction} and \eqref{eq:m2Prediction} to estimate the necessary values of the couplings $\lambda$ and $\Cphi$ to get $m^2=0$,
\begin{align}
    \lambda_{\mathrm{CW}}&\approx -0.31,\\
    \Cphi_{\mathrm{CW}}v^2&\approx -0.21.
\end{align}
Because this parameter point is expected to support a tree-level barrier, and hence $\lambda\sim g^2$, we emphasize that it is only CW-\emph{like}: it does not feature radiative symmetry breaking. We can think of it as "effectively" scale-invariant: the only scale present comes from the cutoff-scale that defines the Wilson coefficient $\Cphi$.

We note this parameter point in figure~\ref{fig:res-cphi-lambda} (marked with $m^2=0$), situated well within the TLB region. We do find viable points in the vicinity of this point, but very few points to the left, which would be likely to have $m^2>0$.

In the above analysis we considered various edge cases that could contradict the assumption that the 3D parameters can be estimated solely from their 4D counterpart, and our conclusion is that we can in fact use this straightforward estimation. For example, to estimate $\lambda_3$ from its 4D counterpart means $\lambda\sim g^{\nlambda} \implies \lambda_3/T\sim g^{\nlambda}$.

\subsection{Identifying first-order phase transitions}\label{ssec:first-order}

With that result in hand, we turn to the question of for which sizes of $\lambda$ should the different power countings be used? It does not seem possible to answer this question within perturbation theory. Instead, we can turn to lattice data to get an appreciation for what sizes of $\lambda$ can yield a first-order phase transition~\cite{Rummukainen:1998as}. It is possible to determine the phase portrait of the theory of~$\mathrm{SU}(2)+$fundamental Higgs in terms of the dimensionless ratio
\begin{equation}
    x\equiv\frac{\lambda_3}{g_3^2}.
\end{equation}
The conclusion is that $x$ must satisfy $x \leq 0.11$ for the theory to feature a first-order phase transition. We can also place a lower bound on this range, from the fact that the lattice studies do not apply when the high-temperature expansion breaks down (and the $\phi^6$ term becomes important). In~\cite{Gould:2019qek}, the lower bound is taken to be $x \geq 0.01$. Putting this together, we have
\begin{equation}\label{eq:su2FOPT}
    0.01 \leq x \leq 0.11
\end{equation} 
for the theory to feature a first-order phase transition from a radiative barrier.

We now wish to translate this bound into regions for our 4D parameters, which determine what power-counting should be used to study that region at high temperature. In our conventions, we have at leading order $x\approx\lambda/2g^2$. Using the numeric value $g = 0.651$, derived using the input parameters as described in Section~\ref{ssec:inputs}, the region in \eqref{eq:su2FOPT} translates to
\begin{equation}
    0.008 \leq \lambda \leq 0.093.
\end{equation}
We will take this region to be $\lambda \sim g^3$. Larger values will not have a radiative barrier and hence will be counted as $\lambda\sim g^2$. For smaller (but still positive) values of $\lambda$, the vector boson becomes too heavy to be described using the high-temperature expansion. This is the region of radiative symmetry breaking, and hence we count this region as $\lambda\sim g^4$. In total, we have
\begin{subequations}
    \label{eq:lambdas}
    \begin{align}
    \lambda \sim g^2:&\quad  \lambda \geq 0.093,  \\
    \lambda \sim g^3:&\quad  0.008 \leq \lambda \leq 0.093, \\
    \lambda \sim g^4:&\quad  0 \leq \lambda \leq 0.008.
\end{align} 
\end{subequations}
For negative values of $\lambda$ we simply mirror this relation, such that, e.g., $\lambda \sim -g^2$ if $ \lambda \leq -0.093$. 

We collate this analysis in figure~\ref{fig:res-cphi-lambda}, where we slice the parameter space of $\lambda$ and $\Cphi$ into regions based on \eqref{eq:lambdas}. The regions represent our prediction for which power-counting should be used for those parameter values, and they correspond to the star-marked regions in figure~\ref{fig:fopt-regions-3d}. The black line corresponds to our simplified prediction of the values necessary to get the correct Higgs mass and vev in \eqref{eq:CphiPrediction}.

Note that this slicing of the parameter space is a somewhat crude approximation. Compare with the sketch of the 3D parameter space in figure~\ref{fig:fopt-regions-3d} which features a rich landscape of different countings from the catalog. Our vertical bands in figure~\ref{fig:res-cphi-lambda} should hence be further sliced into smaller regions. Determining where these regions are mapped to from the 4D parameter space is difficult, because both $m_3^2$ and $\Cphithree$ have strong temperature dependence. This will require a combination of lattice studies and higher order perturbative calculations, and we will leave it for future work. 

%================================================================================================================
\section{Phenomenology}\label{sec:pheno}
%================================================================================================================

In this section, we perform some estimations of whether a first-order phase transition within the SMEFT can give a detectable gravitational wave signal, and if the model can support baryogenesis. 
 As our study is  exploratory, we will here make some simple estimates, leaving detailed calculations for future work---one reason being that the accuracy of GW calculations is still under active exploration, and much evidence has been given that higher order corrections are necessary to accurately determine GW signals~\cite{Croon:2020cgk, Gould:2021oba, Gould:2021dzl, Schicho:2022wty,Ekstedt:2022ceo, Athron:2022jyi, Ekstedt:2022zro, Gould:2023ovu, Gould:2023jbz, Lewicki:2024xan, Chala:2024xll, Gould:2024chm, Ekstedt:2024etx}. 
 
\subsection{Electroweak baryogenesis and sphaleron transitions}
A first-order electroweak phase transition could make way for electroweak baryogenesis~\cite{Morrissey:2012db}, a proposed mechanism that aims to fulfill the three Sakharov conditions~\cite{Sakharov:1967dj} necessary for dynamically generating a non-zero baryon number. 

Sphaleron transitions~\cite{Klinkhamer:1984di}---fluctuations in the electroweak fields which can violate $B-L$ and hence change baryon number---play a key role in electroweak baryogenesis. First, such transitions help generate a nonzero baryon number in the plasma just outside the bubble walls. Second, the sphaleron transitions must be sufficiently suppressed in the broken phase so that they do not wash out the generated baryon number.

In this subsection, we discuss the suppression of the sphaleron rate. We focus on improving the typical leading order measure used to determine if the suppression is large enough for baryogenesis, based on the discussion in~\cite{Patel:2011th}. See~\cite{Balazs:2016yvi, deVries:2017ncy} for discussions of the other factors in electroweak baryogenesis in the SMEFT, and see~\cite{Gan:2017mcv, Phong:2020ybr} for discussions of the sphaleron rate in the same model. For higher order corrections, see~\cite{Arnold:1987mh, Carson:1989rf, Carson:1990jm, Baacke:1993aj, Baacke:1994ix}.\footnote{Note that some of these papers use a different notation for $g_3$ from what we use in this paper.} For a modern discussion on how to solve the sphaleron equations of motion in more generic models, see~\cite{Wu:2023mjb}.

We can formulate an electroweak baryogenesis viability-condition on the sphaleron rate in the broken minimum after the transition by demanding the rate to be suppressed. Typically, this condition is stated as $v_c/T_c\geq 1$, where $v_c$ is the vev (in 4D units) at the critical temperature $T_c$. This condition stems from estimating the sphaleron transition rate as
\begin{equation}
    \Gamma_{\mathrm{sph}}\simeq A e^{-\frac{E_{\mathrm{sph}}}{T}},
\end{equation}
where $E_{\mathrm{sph}}$ is the energy of the sphaleron configuration and $A$ is given by dissipative processes and higher order effects. By estimating that at the critical temperature, $E_{\mathrm{sph}}\sim v_c/g$, and then directly relating this rate to a change in the baryon number, the condition $v_c/T_c\geq 1$ emerges.

As has been emphasized before~\cite{Patel:2011th}, this condition has several problems. One issue is that if care is not taken when calculating $v_c$, gauge dependence will be introduced into the measure, calling its physical relevance into question. As an example, if one calculates $v_c$ and $T_c$ by mixing loop orders in the effective potential, $v_c/T_c$ will be gauge dependent. 

In~\cite{Patel:2011th}, the authors suggest using instead the ratio $\bar{v}/T_c$, where $\bar{v}$ is the minimum in the corresponding dimensionally reduced 3D EFT and $T_c$ is the critical temperature found via an ad-hoc procedure. This strategy does give a gauge invariant measure, but it neglects the fact that the 3D EFT might not have a barrier---as is the case in the $\mathrm{SU(2)}+$Higgs theory. See also the analysis around figure 7 of~\cite{Patel:2011th}.

However, the recent progress in understanding how to model radiative barriers by integrating out intermediate scales, suggests that we could use instead the minimum found from the leading order radiative-barrier potential in \eqref{eq:RBpot}. More generally, we could find the minimum $v_3$ (in 3D units) from any potential in our catalog, and estimate the suppression rate as 
\begin{equation}\label{eq:sphaleronApprox}
    -\log\Gamma_{\mathrm{sph}} \simeq \frac{E_\mathrm{sph}}{T} \sim \frac{v_c}{gT} \sim \frac{v_3}{g_3},
\end{equation}
using the leading order matching relations $\phi_3 = \phi/\sqrt{T}$ and $g_3=g \sqrt{T}$
in Eqs.~(\ref{eq:phi3-matching-LO}) and (\ref{eq:g3-matching-LO}). The measure then becomes
\begin{equation}\label{eq:sphaleronMeasure}
    \frac{v_3}{\sqrt{T}}\geq 1.
\end{equation}
This would be a gauge invariant measure that properly includes the effects of resummation, and which is gauge invariant.\footnote{There is a separate question regarding whether comparing the ratio to $1$ is always a good choice. The required suppression of the rate depends on how much baryon number is generated outside the bubble wall, which can be quite model dependent. See~\cite{Patel:2011th} for further discussion.}

Using the approximation in \eqref{eq:sphaleronApprox}, and our power countings in the catalog in table~\ref{tab:catalog}, we can estimate the sphaleron suppression rate for each case. See the rightmost column in table~\ref{tab:ThermoParams} for a summary of this analysis. Here, we mention that the \sqtlb{} potential (tree-level barrier with a small and negative quartic coupling) shows the most promise among the cases that allowed for a numeric estimate. This case has a relatively large vev just after the transition, $v_3 \sim g^{-1/2} \sqrt{T}$.

This table contains cases that do not allow for a numeric estimate of the sphaleron rate: the RSB cases. The reason is that for the sphaleron to be energetically able to jump over the energy barrier requires both the scalar and gauge fields to be light compared to the thermal fluctuations. But in the RSB case the gauge boson is heavy and hence fluctuations are Boltzmann suppressed. The RSB case is hence an even more natural candidate for electroweak baryogenesis.

\begin{table} 
    \begin{center}
        \begin{tabular}{ c|c|c|c||c }
        Hierarchy & Shorthand & $\alpha/\alpha_{\mathrm{RB}}$ & $\beta/H\Big{/}(\beta/H)_{\mathrm{RB}}$ & $-\log \Gamma_{\mathrm{sph}}$\\
        \hline    
        \multirow{ 3}{*}{$m_3 \sim M \ll \pi T$} & TLB & $1$ & $g^{\frac{3}{2}}$ & $g^{-1}$\\ 
        %\hline
        & SC TLB (1) & $1$ & $g^{\frac{3}{2}}$ & $g^{-1}$ \\
        & SC TLB (2) & $g^{1}$ & $g^{1}$ & $g^{-\frac{1}{2}}$ \\         
        \hline        
        \multirow{ 7}{*}{$m_3 \ll M \ll \pi T$} & \sqtlb{} & $g^{-1}$ & $ g^{\frac{1}{2}}$ & $g^{-\frac{3}{2}}$ \\  
        & SC \sqtlb{} (1) & $g^{-1}$ & $ g^{\frac{1}{2}}$ & $g^{-\frac{3}{2}}$ \\ 
        & SC \sqtlb{} (2) & $1$ & $ g^{\frac{1}{4}}$ & $g^{-\frac{5}{4}}$ \\                       
        & RB & $1$ & $1$ & $g^{-1}$\\ 
        & \rbsix & $1$ & $1$ & $g^{-1}$ \\         
        %\hline
        & SC RB & $g$ & $g^{\frac{1}{4}}$ & $g^{-\frac{1}{2}}$\\ 
        & SC RSB & $ g$ & $ g$  &$g^{-1}$\\                    
        \hline                          
        \multirow{ 2}{*}{$m_3 \ll M \sim \pi T$} & RSB  & $ g^{-2}$ & $ g^{-\frac{1}{2}}$ & --- \\  
        %\hline    
        & \rsbsix & $ g^{-2}$ & $ g^{-\frac{1}{2}}$ & --- \\      
        \end{tabular}
        \end{center}
        \caption{Estimations of the thermodynamic parameters relevant for GW detection and the sphaleron suppression rate, for each entry in the catalog in table~\ref{tab:catalog}.}
        \label{tab:ThermoParams}
\end{table}

\subsection{Gravitational wave signal}\label{ssec:GW}
If a first-order phase transition occurs in the early universe, the collisions of nucleating bubbles can disturb the plasma such that a GW signal is generated. The strength of the signal, and hence its probability of being detected in future experiments, depends mainly on four thermodynamic parameters: the strength of the phase transition $\alpha$, the inverse duration of the phase transition $\beta/H$, the wall velocity $v_w$, and the transition temperature $T_{*}$. This signal can be detected by space-based interferometry experiments~\cite{Kawamura:2011zz, Caprini:2015zlo,Caprini:2019egz, Ruan:2018tsw, TianQin:2015yph}. Unfortunately, as discussed in~\cite{Athron:2023xlk} and shortly recapitulated below, there are several competing definitions for many of these thermodynamic quantities. We will sidestep these issues of ambiguity by making a fairly crude estimate of the GW signal, which will not require detailed accuracy.

In~\cite{Gould:2019qek}, a nonperturbative analysis of the generated GW signal from the $\mathrm{SU}(2)+$ Higgs was performed. By considering this model as a template for models that are SM-like at high temperature---only SM operators and fields, with a radiative barrier from the gauge bosons---the authors concluded that LISA would not be able to detect the GWs generated by such a model. This places doubt on whether a phase transition within the SMEFT could generate such a signal.

However, we have in the catalog in table~\ref{tab:catalog} sketched out several effective potentials that could be realized at high temperatures within the SMEFT. With this enlarged space of possibilities, we wish to estimate if any of the potentials are candidates for generating a detectable GW signal. See also~\cite{Delaunay:2007wb, Huang:2016odd, Cai:2017tmh, Hashino:2022ghd} for other studies of the GW spectrum in models with a $\phi^6$ term.

Calculating these thermodynamic parameters for a particular power-counting is an arduous task. It has been shown that the leading order calculation can often have uncertainties that are as large as several orders of magnitude~\cite{Croon:2020cgk,Gould:2021oba, Lewicki:2024xan}, and hence a higher order calculation is needed to provide more reasonable accuracy. As we are merely looking for candidates, we will instead make a fairly crude estimation, and will make a few simplifying assumptions. The calculation of wall velocity for a given model is still in its infancy (see~\cite{Ai:2023see} for a recent discussion), and hence we will follow~\cite{Gould:2019qek} and set $v_w=1$. 

There have been several candidate temperatures in the literature for the phase transition temperature $T_{*}$. The modern view suggests using the percolation temperature $T_{p}$, which is the temperature at which approximately two-thirds of the observable universe is in the broken phase. For further discussion of the phase transition temperature, see~\cite{Athron:2022mmm, Athron:2023rfq}. Though we expect the correct definition of the phase transition temperature to be important for accurately determining the GW spectrum, the details will not be important for our purposes. We simply note that some potentials of the catalog arise through supercooling, and hence those cases necessitates using a definition of $T_{*}$ that can capture supercooling.\footnote{As is discussed in~\cite{Athron:2022mmm}, using the nucleation temperature---the temperature at which one expects one bubble per Hubble volume---is not sensible in cases with strong supercooling.} 

The phase transition strength $\alpha$ plays a key role in determining the energy budget of GWs in the plasma. Many measures of this strength are discussed in~\cite{Athron:2023xlk}, including using the pressure difference, the trace anomaly, or the pseudotrace anomaly~\cite{Giese:2020rtr, Giese:2020znk}---with the pseudotrace anomaly being the most accurate choice, though a bit more complicated to calculate. Luckily for our case, the difference between the trace anomaly and the pseudotrace anomaly arises beyond leading order~\cite{Tenkanen:2022tly}, and hence we can use the trace anomaly to estimate the size of $\alpha$. The definition of $\alpha$ is then the trace anomaly divided by $3w$, where $w$ is the enthalpy density, which for a bag model equation of state satisfies $w=(4/3)\rho$, where $\rho=T \partial p/\partial T$ is the energy density and $p$ is the pressure. We then have (using $V_4=T V_3$)
\begin{equation}
    \alpha = \left(\frac{\partial p_{\mathrm{sym}}}{\partial T}\right)^{-1}\left(\Delta V_3-\frac{1}{4}\frac{\mathrm{d}\Delta V_3}{\mathrm{d}\log T}\right),
\end{equation}
where $p_{\mathrm{sym}}\sim T^4$ is the pressure in the symmetric phase and $\Delta$ denotes the difference between the symmetric and broken phases: $\Delta X = X_{\mathrm{sym}}-X_{\mathrm{bro}}$.

The inverse duration of the phase transition can be calculated from the temperature dependence of the nucleation rate~\cite{Caprini:2019egz}
\begin{equation} 
    \frac{\beta}{H}=-\frac{\mathrm{d}\log \Gamma}{\mathrm{d}\log T},
\end{equation}
where the nucleation rate is calculated using the framework of thermal escape, described in section~\ref{sec:framework}.

With these definitions in hand, we are ready to estimate the thermodynamic parameters using power counting. As an example, we can consider the case of a radiative barrier. As mentioned above, this example is interesting because Ref.~\cite{Gould:2019qek} demonstrated, using lattice calculations, that SM-like theories with radiative barriers do not have a large enough GW signal to be detected using LISA. The phase transitions are in general not strong enough (too small $\alpha$) and have too short duration (too large $\beta/H$).

Inspired by~\cite{Gould:2019qek}, we note that the 3D EFT describing the radiative barrier only contains temperature dependence through the couplings, and we introduce the shorthand
\begin{equation}
    \eta_X\equiv \frac{\mathrm{d}X}{\mathrm{d}\log T}.
\end{equation}
For a radiative barrier, the quantities needed to evaluate the parameters behave as
\begin{align}
    \Delta V_3 &\sim m_3^2 \phi_3^2 \sim g^3 T^3,\\
    T\frac{\mathrm{d}\Delta V_3}{\mathrm{d}T} &= \frac{1}{2}\eta_{m_3^2} \phi_3^2 -\frac{1}{16 \pi} \eta_{g_3^3} \phi_3^3+\frac{1}{4} \eta_{\lambda_3} \phi_3^4 \sim g^2 T^3,\\
    -\frac{\mathrm{d}\log \Gamma}{\mathrm{d}\log T} &= \int_x\left( \frac{1}{2}\eta_{m_3^2}\phi_3^2 -\frac{1}{16\pi}\eta_{g_3^3} \phi_3^3+\frac{1}{4} \eta_{\lambda_3} \phi_3^4\right)+\mathellipsis\sim g^{-\frac{5}{2}}.
\end{align}
Hence the radiative barrier case has the following parameters,
\begin{align}
    \left.\alpha_{\mathrm{RB}}\right|_{T_{*}}&\sim g^2,\\
    \left.\left(\frac{\beta}{H}\right)_{\mathrm{RB}}\right|_{T_{*}}&\sim g^{-\frac{5}{2}}.
\end{align}
These estimations offer a reference point. We have performed the same analysis as above (with some complications explained toward the end of this subsection) for all the power-counting schemes in the catalog, and present the result in table~\ref{tab:ThermoParams}. In that table, we compare the estimated parameters to that of the radiative barrier case. From the table, we conclude that the most promising candidate is that of \sqtlb{} (and its supercooled variants), which corresponds to a tree-level barrier with a small and negative quartic coupling $\lambda_3$. Indeed, \sqtlb{} is the only family of cases which seems to have a larger $\alpha$ and smaller $\beta/H$ than the radiative-barrier case.

However, by inspecting the "blob" in figure 2 of~\cite{Gould:2019qek}, we conclude that the blob is quite large. And because our power-counting estimate is rather crude, if a given power-counting in fact moves this blob in one of the two promising directions, it might be enough for the blob to somewhat overlap the sensitivity region. With this widened condition, also the TLB power-counting scheme is interesting.

Finally, we note some complications that can arise when calculating the thermodynamical parameters. First, our potentials with radiative symmetry breaking explicitly depend on $T$, and hence the analysis cannot be performed purely in terms of the $\eta$'s introduced above. We also note that supercooled potentials require a little extra care. That is because evaluating $\alpha$ requires evaluating the potential in the broken minimum, which is not accessible using the supercooled version of a given potential. Instead, one can use the supercooled potential to find the percolation temperature, and then use the original potential (from which it supercooled) to find the broken minimum and hence $\alpha$. This approach was pioneered by Ref.~\cite{Kierkla:2023von} and is further explained therein.

\subsection{Higgs pair production}

Let us finally briefly comment on implications for Higgs pair production at the LHC. This process is intimately connected to the physics of the electroweak phase transition, as it involves the Higgs quartic coupling present in the Higgs potential. In the SM there are two one-loop diagrams that contribute at leading order, one involving both the Higgs-top quark coupling and the triple Higgs coupling $\lambda_{HHH}$, which is in turn proportional to the quartic coupling parameter $\lambda$, the other involving two powers of the Higgs-top coupling. There is destructive interference between the two diagrams and therefore a rather small cross-section. In the SMEFT, both diagrams receive corrections from dimension-six operators, and in addition there are several new diagrams involving SMEFT operators both at tree level and one-loop level. In particular the triple Higgs coupling in the SMEFT involves the quartic $\lambda$ as well as the three Wilson coefficients $\Cphi$, $\Cphibox$ and $\CphiD$. As we have seen, for scenarios with a first-order phase transition, $\lambda$ can have values substantially different from the SM value, as well as the opposite sign from the SM. Moreover, the triple Higgs vertex in the SMEFT is momentum-dependent, proportional to $\Cphibox$ and $\CphiD$. Precise measurements of the differential cross-section and its momentum dependence at the high-luminosity LHC could therefore test the scenarios presented here. A detailed study of Higgs pair production is beyond the scope of this paper, and we refer the reader to our previous paper~\cite{Camargo-Molina:2021zgz} for more details, and, e.g., \cite{Alasfar:2023xpc} for further discussion.\footnote{We note, however, that most sources discussing Higgs pair production in the SMEFT choose an input scheme that fixes the Higgs quartic coupling to the same value as in the SM, making the analysis of the potential more complicated.}

%================================================================================================================
\section{Discussion}\label{sec:discussion}
%================================================================================================================
In contrast to the SM, the SMEFT can have a first-order electroweak phase transition and satisfy experimental constraints. There are, simultaneously, reasons to be skeptical and reasons to believe such a scenario is realized in nature. We will repeat these reasons here and describe why we are somewhat optimistic.

The reasons to be skeptical include the necessity of fairly large Wilson coefficients, which signals that the physics encoded by the higher dimensional operators live at a scale rather close to the electroweak scale,
\begin{equation}
    \Cphi\sim \frac{\lambda}{v^2}\sim \frac{1}{\Lambda^2}\implies \Lambda \sim \frac{v}{\sqrt{\lambda}}\gtrsim  \frac{v}{g},
\end{equation}
with the last approximate inequality assuming $\lambda \sim g^2$ (which is one of the cases we consider). This issue was emphasized already in Refs.~\cite{Damgaard:2015con} and \cite{Postma:2020toi}: it is very hard to find realistic models with large enough Wilson coefficients for a barrier to form in the scalar potential, yet not yield any large differences to collider physics such that they have not been observed. By considering such large Wilson coefficients, are we not taking the principles of Wilsonian EFTs seriously enough? 

Furthermore, the estimates we have performed in this paper do not include higher order effects, and there are few nonperturbative studies for us to measure against. It is possible that some of the cases we consider will turn out to not be perturbatively viable.

On the other hand, here are a few reasons to be more optimistic. Returning to Wheeler's principle of \emph{radical conservatism}~\cite{Thorne:2019scz}, we can take the absence of evidence for BSM fields dynamic at the electroweak scale as evidence of their absence. If we simultaneously accept the idea of electroweak baryogenesis---or at the very least the need to suppress sphaleron transitions after the phase transition---we are faced with the task of finding a barrier in the potential without including BSM fields. This reasoning leads us directly to seriously consider how a barrier can arise in the SMEFT, overriding the concerns caused by unnaturally large values of the Wilson coefficients.

After all, the Wilsonian perspective on the values of the parameters in the scalar potential already woefully fails to predict the SM values. If we do not have natural values for the parameters of dimension zero and dimension two, why should they be natural for dimension six? Perhaps using naturalness arguments to predict their sizes has been misguided all along. Perhaps it would be a mistake to do the same for the dimension-six parameter $\Cphi$. 

Taking this reasoning seriously, the crude results of table~\ref{tab:ThermoParams} suggest that we should pay extra attention to the cases that particularly suppress sphaleron rates: the RSB (radiative symmetry breaking) potential and the \sqtlb{} potential (tree-level barrier with small and negative quartic coupling). \sqtlb{} is also the potential we judge most likely to have a strong enough GW signal to be detected. Starting with our stated principles and following the path of reason to the end has thus raised our credence in the possibility of detecting a GW signal from the SMEFT.

That said, even if new data is released tomorrow which eliminates the possibility of a first-order phase transition within the SMEFT, we still consider the construction and analysis of the catalog in table~\ref{tab:catalog} to be useful beyond its relevance for the SMEFT. It serves as a didactic demonstration on how to use reasoning based on hierarchies of scale and power counting to establish a well-behaved leading order potential with a barrier. The inclusion of the $\phi^6$ term offered a number of interesting challenges, and the exploration of supercooling in power-counting terms emphasized the emergence of new scales in-between the soft and ultrasoft scales, as discussed in appendix~\ref{app:pc}. These scales add to the already rich hierarchy presented in~\cite{Gould:2023ovu}.

We also emphasize that the construction of the catalog does not cover all possible EFT extensions of the SM. For example, the Higgs Effective Field Theory (HEFT) is a superset of the SMEFT, and additionally covers, e.g., cases with BSM fields that decouple only after EWSB. The electroweak phase transition within the HEFT has been studied in e.g.~\cite{Banta:2022rwg, Alonso:2023jsi}, though as of yet there are no such studies that use the modern methods of dimensional reduction. Their perturbative accuracy is hence not yet well established. We encourage interested parties to explore what a dimensionally reduced HEFT would entail. Note that the construction of the HEFT relies on the technique of functional matching~\cite{Cohen:2020fcu}---the same technique used to construct e.g. the radiative symmetry breaking potential. Performing a functional matching from the HEFT to a 3D theory which includes the pertinent operators could perhaps be a fruitful approach.

It is important to comment on the expected impact of NLO corrections on our findings. As we note in Secs.~\ref{ssec:first-order} and \ref{ssec:GW}, our analysis represents a leading-order framework, and we have emphasized that higher-order corrections are necessary for precision predictions. We anticipate that NLO corrections will refine, rather than invalidate, the qualitative classification presented in our catalog. Their primary effect would be to quantitatively shift the boundaries between the phenomenological regions depicted in figure \ref{fig:res-cphi-lambda}, i.e.~the sharp vertical lines, which represent our LO estimates based on the analysis in Sec.~\ref{ssec:first-order}, would be replaced by theoretical uncertainty bands. However, the existence of these distinct scenarios, i.e. tree-level barriers, radiative barriers, is a direct consequence of the features of the potential which we identify using our power-counting framework. Our work provides a map of the SMEFT landscape, which is a necessary prerequisite for targeted NLO studies in the future.

For example, lattice studies can be performed, to better understand the phase portrait of $\mathrm{SU}(2)+\phi^6+$Higgs. Such studies would improve our understanding of the parameter regions of the SMEFT that describe first-order phase transitions. Furthermore, the various perturbative prescriptions outlined in this paper can be pushed to higher orders, as has been done for the radiative barrier case~\cite{Ekstedt:2024etx}. This would improve our understanding of the convergence of these prescriptions, and would increase the robustness of the GW signal calculations against their inherent large uncertainties---refining the accuracy of our leading order estimates. The catalog could also be used as a stepping-off point to study the electroweak phase transition in models with a $\phi^6$ term and new fields.

Finally, we find the case of a tree-level barrier with a small and negative $\lambda_3$ (\sqtlb) to be particularly interesting, and encourage future studies of the SMEFT in the corner of parameter space that realizes that potential.

\section*{Acknowledgments}

\noindent{}We thank Andreas Ekstedt, Oliver Gould, and Tuomas Tenkanen for enlightening discussions. Part of this work was inspired from discussions held at the workshop "Thermal Field Theory and Early Universe Phenomenology at the Botanical Garden" held at the Swedish Collegium for Advanced Study in June 2023. JL's work was funded by E och K.G. Lennanders stipendiestiftelse. The computations were performed on the Google Cloud Platform under the research credits program. The figures were made with the \texttt{Julia} package \texttt{Makie}~\cite{DanischKrumbiegel2021}.

\appendix

\section{Simplified analysis of Higgs mass and vacuum expectation value}
In this section we perform a simplified analysis of calculating the Higgs mass and vev, using the 1-loop effective potential.  We do not use this calculation when performing the scan, as we judge the assumptions it is based on (Higgs being lighter than the electroweak gauge bosons and the top quark), to not apply. However, the analysis serves as a guide for which SMEFT parameters we expect to be relevant.

As such, we will add 1-loop corrections to the classical potential in \eqref{eq:classical-potentialSMEFT}. The 1-loop contributions to the effective potential in $\overline{\mathrm{MS}}$, with $\mu$ the renormalization scale, and with finite-temperature corrections included, can be written~\cite{Coleman:1973jx, Laine:2016hma}
\begin{align}
    V_1(\phi)&=\sum_a n_a I_a(M_a^2,T),\\
    I_a(M_a^2,T)&=J_a(M_a^2)+\frac{T^4}{2\pi^2}J_a^{T\neq 0}(\frac{M_a^2}{T^2}),\\
    J_a(x)&=\frac{x^2}{64\pi^2}\left(\log\frac{x}{\mu^2}-C_a\right),\quad
    C_a=
    \begin{cases}
        \frac{3}{2} & \text{for scalars and fermions,}\\
        \frac{5}{6} & \text{for gauge bosons,}\\
      \end{cases}\label{eq:CWIntegralFunc}\\
      J_a^{T\neq 0}(x)&=\int_{0}^{\infty}\diff{y} y^2\log\left(1\pm e^{-\sqrt{y^2+x}}\right), \quad \begin{cases}
      + & \text{for fermions,}\\
      - & \text{for bosons.}\\
    \end{cases}\label{eq:finTIntegralFunc}
\end{align}
Here $J_a(x)$ is the zero-temperature contribution (often dubbed the Coleman-Weinberg potential), $n_a$ denotes the number of degrees of freedom of the field $a$ with field-dependent mass $M_a$, including a sign for the fermion loop.

To find the 1-loop corrected minimum, we will consider the leading order effective potential at zero temperature to be given by integrating out the gauge bosons and the top quark. This gives the potential
\begin{align}
    V_{\mathrm{LO}}(\phi)&=V_0(\phi)+n_W J_W (M^2_W)+n_Z J_Z(M^2_Z)+n_t J_t (M^2_t),\\
    n_Z &= 3, \quad M_Z^2=\frac{g^2+g'^2}{4}\phi^2\left(1+\frac{1}{2}\CphiD\phi^2\right),\\
    n_W &= 6, \quad M_W^2=\frac{g^2}{4}\phi^2,\\    
    n_t &= -12, \quad M_t^2=\frac{\yt^2}{2}\phi^2.
\end{align}
This leading order potential is gauge invariant because the unphysical modes give nonzero contributions first at next to leading order~\cite{Metaxas:1995ab}. We fix the renormalization scale to be the same as the measured Z mass, $\mu=\left.M_Z\right|_{\phi=v}$. Then we can find an expression for the vacuum expectation value $v$ by minimizing this potential,
\begin{equation}\label{eq:vevLO}
    \left.\partial_\phi V_{\mathrm{LO}}\right|_v = 0.
\end{equation}

We also calculate the squared mass of the Higgs field from the LO potential,
\begin{align}
    m_H^2&=-m^2+\frac{3 \lambda  v^2}{2}-2C^{\text{kin}}v^2\left(\frac{3 \lambda  v^2}{2}-m^2\right)-\frac{15}{4} \Cphi v^4+\Delta m_H^2,\label{eq:mH2LO}\\
    C^{\text{kin}}&=\frac{\CphiD}{4}-\Cphibox.
\end{align}

Because we want to make an estimation of the allowed parameter space, we will make some simplifying assumptions. Starting with the expression for the vev, we can neglect $\CphiD$ to get an estimate of the correction from the 1-loop terms. Then we find that the vacuum expectation value must approximately satisfy
\begin{equation}\label{eq:vevEqCorrected}
    0\approx-m^2+\frac{1}{2}(\lambda+\delta)v^2-\frac{3\Cphi}{4}v^4,
\end{equation}
where $\delta\approx -0.008$ is given by
\begin{align}
    \delta=\frac{1}{8\pi^2}&\left(6\left(\frac{g^2}{4}\right)^2\left(\log\frac{g^2}{g^2+g'^2}-\frac{1}{3}\right)-\left(\frac{g^2+g'^2}{4}\right)^2\right.\\
    &-12\left(\frac{y_t^2}{2}\right)^2\left(\log \frac{2y_t^2/2}{\left(g^2+g'^2\right)/4}-1\right)\biggr).\nonumber
\end{align}

Neglecting the dimension-six Wilson coefficients' contribution to the 1-loop correction of the mass, gives $\Delta m_H^2\approx -0.04 v^2$, where 
\begin{align}
    \Delta m_H^2\approx \partial^2_\phi V_1 =\frac{3v^2}{16 \pi^2}&\left(6\left(\frac{g^2}{4}\right)^2\left(\log\frac{g^2}{g^2+g'^2}+\frac{1}{3}\right)+\left(\frac{g^2+g'^2}{4}\right)^2\right.\label{eq:mH2EqCorrected}\\
    &-12\left(\frac{y_t^2}{2}\right)^2\left(\log \frac{y_t^2/2}{\left(g^2+g'^2\right)/4}-\frac{1}{3}\right)\biggr)\nonumber.
\end{align}
Using the approximate constraints set by \eqref{eq:vevEqCorrected} and \eqref{eq:mH2EqCorrected}, and neglecting the effects of $\CphiD$ and $\Cphibox$, we can solve these equations for $m^2$ and $\Cphi$ in terms of the measured Higgs mass squared, $M_H^2$, the vacuum expectation value, $v$, and $\lambda$. The result is a linear relation between $\Cphi$ and $\lambda$,
\begin{align}
    \Cphi &\approx \frac{1}{3 v^2}\left(\lambda -\frac{\delta}{2}-\frac{M_H^2-\Delta m_H^2}{v^2}\right),\label{eq:CphiPrediction}\\
    m^2 &\approx \frac{v^2}{4}\left(\lambda+\frac{5 \delta}{2}+\frac{M_H^2-\Delta m_H^2}{v^2}\right).\label{eq:m2Prediction}
\end{align} 

Formally, it is only allowed to include these loop corrections for small values of $\lambda$. But we will use the loop-corrected formulas of \eqref{eq:vevEqCorrected} and \eqref{eq:mH2EqCorrected} for the whole parameter space anyway. The logic is that the modifications they introduce are negligible in those regions where they are formally forbidden, but important in the regions where they are formally allowed.

\section{Estimating coefficients with cancellations}\label{app:estimating-coeffs}
Recall that the scalar squared mass in the 3D theory can be smaller than originally expected due to a cancellation between the leading terms,
\begin{equation} 
    m_3^2\sim m^2+\# g^2 T^2\sim g^n T^2,
\end{equation}
with $n>2$. This cancellation is necessary for a radiative barrier to be able to form, as seen in the catalog in table~\ref{tab:catalog}. This requirement is not a real cause for concern, because for a radiative barrier $m_3^2$ has the strongest temperature dependence. Hence, there is no conflict between requiring the terms in $m_3^2$ to cancel, and the rest of the analysis.

Could a similar cancellation also occur for $\lambda_3$? The leading order relation for $\lambda_3$ is
\begin{equation}
    \lambda_3/T \sim \frac{\lambda}{2}-\Cphithree,
\end{equation}
where we replaced $-\Cphi T^2$ with $\Cphithree$, which holds at leading order. Assuming that $\Cphithree$ is never expected to be larger (in a power-counting sense) than $\lambda$, we can hence use the size of $\lambda$ to estimate $\lambda_3/T$. Unless there is a cancellation between the two terms, of course.

For such a cancellation to take place, we require $\lambda<0$ and $\Cphithree\sim \lambda$. There is no inherent contradiction in this requirement, but it turns out to be problematic perturbatively. Let us call the temperature at which this cancellation occurs $T_{\mathrm{cancel}}$,
\begin{equation}
    \left.\lambda_3/T\right|_{T_{\mathrm{cancel}}}\sim g^{\nlambda}, \qquad\nlambda>\nC.
\end{equation}
Inspecting the catalog in table~\ref{tab:catalog} we find that there are no items satisfying this requirement. Indeed, it seems as requiring $\Cphithree$ to be large enough to cancel against $\lambda$ simultaneously makes the $\phi_3^6$ term too large for perturbativity to apply.

Beyond this problem, there is a question of fine-tuning. Because $\Cphithree$ depends on the temperature as $T^2$, it has the same sensitivity to the temperature as $m_3^2$. If we simultaneously require the temperature to be close to $T_{\mathrm{cancel}}$, and to be close to $T_c$ (determined by a relation such as in \eqref{eq:critMassRB6}), some very precise fine-tuning must occur. This is of course possible but is expected to be rare in the wider parameter space of the SMEFT.

We hence simply note the possible cancellation in $\lambda_3$ as a curiosity, but do not study it further.
%================================================================================================================
\section{Notes on power counting}\label{app:pc}
%================================================================================================================
In this appendix, we give more details on the derivations of the different power countings in the catalog of table~\ref{tab:catalog}. As a reminder, we are assuming that the different components scale with the power-counting parameter $g$ as
\begin{align}
    m_3^2 &\sim g^{n_{m^2}} T^2,\label{eq:nm2App}\\
    \lambda_3 &\sim g^{n_\lambda} T,\label{eq:nlambdaApp}\\
    \phi_3 &\sim g^{n_\phi} \sqrt{T},\label{eq:nphiApp}\\    
    \Cphithree &\sim g^{n_C},\label{eq:ncApp}
\end{align}
and $g_3\sim g \sqrt{T}$. Note that there are four perturbativity constraints that may apply to any given power counting,
\begin{enumerate}
    \item[\textbf{A}] $m_3^2 \phi_3^2 \gg g^6 T^3$ (Leading order perturbative),
    \item[\textbf{B}] $m_3^2 \ll T^2$ (High temperature expansion applies to Higgs),
    \item[(\textbf{C})] $m_3^2 \ll g_3^2 \phi_3^2$ (Spatial gauge boson can be integrated out in 3D EFT),
    \item[(\textbf{D})] $g_3^2 \phi_3^2 \ll T^2$ (High temperature expansion applies to gauge boson). 
\end{enumerate}
The constraints \textbf{A} and \textbf{B} are hard constraints that must be obeyed for all the power-counting schemes, otherwise the perturbative description will not be useful. But the constraints (\textbf{C}) and (\textbf{D}) are more context-dependent.

Constraint \textbf{A} ensures that our theory contains contributions to the free energy that are parametrically larger than the contribution from the ultrasoft scale $p\sim g^2 T$. The physics at this scale---the spatial modes of the gauge bosons in the symmetric minimum---becomes strongly coupled and is nonperturbative~\cite{Linde:1980ts}. For our perturbative calculations of the free energy to yield meaningful predictions, we need the leading order to be dominant to these contributions, which are of the order $g_3^6\sim g^6 T^3$. We pick $m_3^2 \phi_3^2$ as a representative of the leading order free energy, as it is present in all the cases we consider.

Constraint \textbf{B} ensures that the high temperature expansion applies to the scalar fields (Higgs and Goldstones). As we described in section~\ref{sec:framework}, this is needed for thermal escape, and allows us to use the framework of classical bubble nucleation to calculate the nucleation rates.

Constraint (\textbf{C}) ensures that there is a hierarchy of scales between the scalars and the gauge boson modes. This means a radiative barrier can form and that it can be described perturbatively. This constraint is only relevant for radiative barriers, where the gauge boson is integrated out.

Constraint (\textbf{D}) ensures that the high temperature expansion applies to the gauge boson modes. This is not a hard constraint, and we will see examples below of power countings which violate it but are still viable perturbative descriptions.

When we insert our assumed scaling laws for our parameters (\Crefrange{eq:nm2App}{eq:ncApp}), the perturbativity constraints yield
\begin{enumerate}
    \item[\textbf{A}] $n_{m^2}<6-2n_\phi$,
    \item[\textbf{B}] $n_{m^2}>0$,
    \item[(\textbf{C})] $n_{m^2} > 2 + 2n_\phi$,
    \item[(\textbf{D})] $1+n_\phi > 0$. 
\end{enumerate}

With these constraints in hand, we are now ready to start exploring the perturbative viability of different potentials. To do so, we will start playing around with including different terms, and see what conditions we find.
\subsection{Tree-level terms only}
Let us by start with assuming that all loop corrections are suppressed and that the tree-level potential is an accurate description. Then the potential is
\begin{equation}\label{eq:TLBAppendix}
    V_3(\phi_3)=\frac{1}{2}m_3^2 \phi_3^2+\frac{1}{4}\lambda_3 \phi_3^4+\frac{1}{6}\Cphithree \phi_3^6.
\end{equation}
Balancing the terms---and assuming that the 1-loop gauge boson contribution is negligible---gives
\begin{equation}
    m_3^2 \phi_3^2\sim\lambda_3 \phi_3^4\sim\Cphithree \phi_3^6 \gg g_3^3 \phi_3^3.
\end{equation}
From which we deduce the relations
\begin{align}
    \nmSq &= \nlambda + 2\nphi,\\
    \nC &= \nlambda - 2\nphi,\\
    \nlambda &< 3-\nphi.\label{eq:nlConstraintTLB}
\end{align}
With these expressions we can use the perturbativity constraints \textbf{A, B} and \textbf{(D)}, and the constraint from \eqref{eq:nlConstraintTLB}, to constrain $\nphi$,
\begin{align}
    \mathbf{A} &\implies \nphi < \frac{1}{4}\left(6-\nlambda\right),\\
    \mathbf{B} &\implies \nphi > -\frac{1}{2}\nlambda,\\    
    \mathbf{(D)} &\implies \nphi > -1,\\   
    \mathrm{\eqref{eq:nlConstraintTLB}}  &\implies \nphi < 3-\nlambda.     
\end{align}

These constraints are still fairly loose. There are many solutions among $\nlambda$ and $\nphi$ for which they are satisfied. To achieve some regularity, we will assume that $\nlambda\geq2$ and that $\nlambda$ only takes integer values. Then we can consider each case one by one.
\begin{itemize}
    \item $\nlambda=2$:
        \begin{equation}
            -1<\nphi<1,
        \end{equation}
        from which we can pick $\nphi=0$ as a representative. This is the case "TLB" in the catalog, and the counting is
        \begin{equation}
            m_3^2 \sim g^2 T^2,\quad \lambda_3 \sim g^2 T,\quad \phi_3 \sim \sqrt{T}, \quad \Cphithree \sim g^2.
        \end{equation}
    \item $\nlambda=3$:
        \begin{equation}
            -1<\nphi<0,
        \end{equation}
        from which we can pick $\nphi=-1/2$. This is the case "\sqtlb" in the catalog, and the counting is
        \begin{equation}
            m_3^2 \sim g^2 T^2,\quad \lambda_3 \sim g^3 T,\quad \phi_3 \sim g^{-\frac{1}{2}}\sqrt{T}, \quad \Cphithree \sim g^4.
        \end{equation}
        \item $\nlambda=4$:
        \begin{equation}
            -1 < \nphi < -1,
        \end{equation}
        with no possible solutions---there is no possible perturbative realization of the potential in \eqref{eq:TLBAppendix} with this scaling.      
\end{itemize}
\subsection{Supercooled tree-level barriers}
We can lightly modify the tree-level barrier case to also model supercooled phase transitions, by repeating the steps but instead assuming that the $\phi_3^6$ term is negligible.
 
The potential is
\begin{equation}\label{eq:SCTLBAppendix}
    V_3(\phi_3)=\frac{1}{2}m_3^2 \phi_3^2+\frac{1}{4}\lambda_3 \phi_3^4.
\end{equation}
Balancing the terms---assuming that the $\phi_3^6$ term and the 1-loop gauge boson contribution are negligible---gives
\begin{equation}
    m_3^2 \phi_3^2\sim\lambda_3 \phi_3^4\gg\Cphithree \phi_3^6,~g_3^3 \phi_3^3.
\end{equation}
From which we deduce the relations
\begin{align}
    \nmSq &= \nlambda + 2\nphi,\\
    \nC &> \nlambda - 2\nphi,\label{eq:ncConstraintSCTLB}\\
    \nlambda &< 3-\nphi.\label{eq:nlConstraintSCTLB}
\end{align}
With these expressions we can use the perturbativity constraints \textbf{A, B} and \textbf{(D)}, and the constraint from \eqref{eq:nlConstraintSCTLB}, to constrain $\nphi$,
\begin{align}
    \mathbf{A} &\implies \nphi < \frac{1}{4}\left(6-\nlambda\right),\\
    \mathbf{B} &\implies \nphi > -\frac{1}{2}\nlambda,\\    
    \mathbf{(D)} &\implies \nphi > -1,\\   
    \mathrm{\eqref{eq:nlConstraintSCTLB}}  &\implies \nphi < 3-\nlambda.     
\end{align}

The simple difference between the regular TLB case and the supercooled variant considered here, is in the strict equality of \eqref{eq:ncConstraintSCTLB}. This opens up the possibility space further. First of all, it is possible to have a larger $\nC$ than for TLB and \sqtlb, and the rest of the parameters the same. This automatically satisfies the power counting restrictions, and will have the same properties in general as the regular potentials. We label these supercooled cases "SC TLB (1)", with SC short for supercooled, and similarly for \sqtlb.

The second option is to assume the same $\nC$ as for TLB and \sqtlb, and seeing what else can change. We will again assume that $\nlambda\geq2$ and that $\nlambda$ only takes integer values. The result is then,
\begin{itemize}
    \item $\nlambda=2$:
        \begin{equation}
            0<\nphi<1,
        \end{equation}
        from which we can pick $\nphi=1/2$ as a representative. This is what we label as "SC TLB (2)" in the catalog, and the counting is
        \begin{equation}
            m_3^2 \sim g^{3} T^2,\quad \lambda_3 \sim g^2 T,\quad \phi_3 \sim g^{\frac{1}{2}}\sqrt{T}, \quad \Cphithree \sim g^2.
        \end{equation}
    \item $\nlambda=3$:
        \begin{equation}
            -1/2<\nphi<0,
        \end{equation}
        from which we can pick $\nphi=-1/4$. This is the case "SC \sqtlb{} (2)" in the catalog, and the counting is
        \begin{equation}\label{eq:scalingscsqtlb}
            m_3^2 \sim g^{\frac{5}{2}} T^2,\quad \lambda_3 \sim g^3 T,\quad \phi_3 \sim g^{-\frac{1}{4}}\sqrt{T}, \quad \Cphithree \sim g^4.
        \end{equation}
        \item $\nlambda=4$:
        \begin{equation}
            -1 < \nphi < -1,
        \end{equation}
        with no possible solutions---there is no possible perturbative realization of the potential in \eqref{eq:SCTLBAppendix} with this scaling.      
\end{itemize}
We include these cases in the catalog in table~\ref{tab:catalog}. Note the presence of a new scale, $m_3\sim g^{5/4}T$, situated in between the soft and supersoft scales in the hierarchy in \eqref{eq:scale-hierarchy}.

\subsection{Radiative barrier}\label{ssec:RBApp}
Here we will assume that there is a hierarchy between the scalar fields and the gauge boson, which allows us to include a $\phi_3^3$ term in the potential:
\begin{equation}\label{eq:RBAppendix}
    V_3(\phi_3)=\frac{1}{2}m_3^2 \phi_3^2-\frac{g_3^3}{16\pi}\phi_3^3+\frac{1}{4}\lambda_3 \phi_3^4+\frac{1}{6}\Cphithree \phi_3^6.
\end{equation}
Balancing the terms gives
\begin{equation}
    m_3^2 \phi_3^2\sim\lambda_3 \phi_3^4\sim g_3^3 \phi_3^3\sim\Cphithree \phi_3^6.
\end{equation}
From which we deduce the relations 
\begin{align}
    \nmSq &= 3 + \nphi,\\
    \nlambda &= 3 - \nphi,\\
    \nC &= 3 - 3\nphi.
\end{align}
With these expressions we can use the perturbativity constraints \textbf{A, B, (C)} and \textbf{(D)} to constrain $\nphi$,
\begin{align}
    \mathbf{A} &\implies \nphi < 1,\\
    \mathbf{B} &\implies \nphi > -3,\\    
    \mathbf{(C)} &\implies \nphi < 1,\\   
    \mathbf{(D)} &\implies \nphi > -1.       
\end{align}
These constraints already resolve to 
\begin{equation}
    -1 < \nphi < 1,
\end{equation}
from which we can pick the representative $\nphi = 0$. This yields the counting
\begin{equation}
    m_3^2 \sim g^3 T^2,\quad \lambda_3 \sim g^3 T,\quad \phi_3 \sim \sqrt{T}, \quad \Cphithree \sim g^3.
\end{equation}
This counting is analogous to the simple example we studied in section~\ref{sec:catalog}, with the caveat that here also the $\phi^6$ term is included. But because the $\phi^6$ coefficient did not play a role in the derivation above, we can also use the same counting but with $\nC > 3$ to represent potentials with radiative barriers but no $\phi^6$ term. Hence, we have here derived the countings named "RB" and "\rbsix" in the catalog.

We also note that if the $\phi^6$ term is present, the $\phi^4$ term is not needed to stabilize the potential. Furthermore, because the coefficient $\nlambda$ did not play a role in the derivation, it is possible to consider $\nlambda>3$. This allows us to drop the $\phi^4$ term from the potential. We choose to still label this scenario as \rbsix, see figure~\ref{fig:fopt-regions-3d}.
\subsection{Radiative symmetry breaking}\label{ssec:RSBApp}
Here we will assume that there is no hierarchy between the gauge boson and the thermal fluctuations, and that:
\begin{equation}\label{eq:RSBAppendix}
    V_3(\phi_3)=\frac{1}{2}m_3^2 \phi_3^2+\frac{9}{T}J_{A}(g_3^2 \phi_3^2)+9 T^3 J^{T\neq 0}_{A}(\frac{g_3^2 \phi_3^2}{T^2})+\frac{1}{4}\lambda_3 \phi_3^4+\frac{1}{6}\Cphithree \phi_3^6.
\end{equation}
See \eqref{eq:CWIntegralFunc} and \eqref{eq:finTIntegralFunc} for definitions of the loop functions. Balancing the terms gives
\begin{equation}
    m_3^2 \phi_3^2\sim\lambda_3 \phi_3^4\sim \frac{g_3^4 \phi_3^4}{T}\sim T^3 \sim\Cphithree \phi_3^6.
\end{equation}
From which we deduce the relations
\begin{align}
    \nmSq &= 2,\\
    \nlambda &= 4,\\
    \nC &= 6,\\
    \nphi &= -1.
\end{align}
As these relations completely determine our scaling parameters, we simply need to make sure that the perturbativity constraints are obeyed. The relevant constraints are \textbf{A, B} and \textbf{(C)},
\begin{align}
    \mathbf{A}:\quad \nmSq = 2 &<8 \quad(\text{OK!}),\\
    \mathbf{B}:\quad \nmSq = 2 &> 0 \quad(\text{OK!}),\\    
    \mathbf{(C)}:\quad \nmSq = 2 &> 0 \quad(\text{OK!}).      
\end{align}
The counting is 
\begin{equation}
    m_3^2 \sim g^2 T^2,\quad \lambda_3 \sim g^4 T,\quad \phi_3 \sim g^{-1}\sqrt{T}, \quad \Cphithree \sim g^6.
\end{equation}
This counting models radiative symmetry breaking. Because the $\phi^6$ coefficient did not play a role in the derivation above, we can also use the same counting but with $\nC > 6$ to represent potentials with radiative symmetry breaking but no $\phi^6$ term. Hence, we have here derived the countings named "RSB" and "\rsbsix" in the catalog.

We would also like to note that the $\phi^4$ coefficient also did not play a role in the derivation above. It is hence possible to consider $\nlambda >4$, dropping the $\lambda_3 \phi_3^4$ term from the potential. We choose to still label this scenario with RSB and \rsbsix. See figure~\ref{fig:fopt-regions-3d}.

\subsection{Supercooled radiative barrier and  radiative symmetry breaking}
We can lightly modify the radiative barrier example in subsection~\ref{ssec:RBApp} to also model supercooled phase transitions, by repeating the steps but instead assuming that the $\phi_3^4$ and $\phi_3^6$ terms are negligible,
\begin{equation}\label{eq:SCRBAppendix}
    V_3(\phi_3)=\frac{1}{2}m_3^2 \phi_3^2-\frac{g_3^3}{16\pi}\phi_3^3\gg\frac{1}{4}\lambda_3 \phi_3^4,~\frac{1}{6}\Cphithree \phi_3^6.
\end{equation}
Balancing the terms gives the relations
\begin{align}
    \nmSq &= 3 + \nphi,\\
    \nlambda &> 3 - \nphi,\label{eq:scrbconstraintApp}\\
    \nC &> 3 - 3\nphi.
\end{align}
The perturbativity constraints and the constraint from \eqref{eq:scrbconstraintApp} give
\begin{equation}
    -1 < \nphi < 1, \quad \nphi >3-\nlambda,
\end{equation}
This set of constraints is still quite loose. Just as for the tree-level potential, we can explore it by assuming $\nlambda\geq 2$ and only considering integer powers for $\nlambda$.
\begin{itemize}
    \item $\nlambda=2$:
        \begin{equation}
            1<\nphi<1,
        \end{equation}
        with no possible solutions---this power-counting is not perturbatively viable.
    \item $\nlambda=3$:
        \begin{equation}
            0<\nphi<1,
        \end{equation}
        from which we can pick $\nphi=1/2$. This is the case of "supercooled RB" in the catalog, and the counting is\footnote{The mass scale $\left(g^{\frac{7}{2}}T^2\right)^{\frac{1}{2}}=g^{\frac{7}{4}}T$ lies between the supersoft and the ultrasoft scales---see \eqref{eq:scale-hierarchy}.}
        \begin{equation}
            m_3^2 \sim g^{\frac{7}{2}} T^2,\quad \lambda_3 \sim g^3 T,\quad \phi_3 \sim g^{\frac{1}{2}}\sqrt{T}, \quad \Cphithree < g^{\frac{3}{2}}.
        \end{equation}
        \item $\nlambda=4$:
        \begin{equation}
            -1 < \nphi < 1,
        \end{equation}
        from which we can pick $\nphi=0$. This is the case of "supercooled RSB" in the catalog, and the counting is
        \begin{equation}
            m_3^2 \sim g^{3} T^2,\quad \lambda_3 \sim g^4 T,\quad \phi_3 \sim \sqrt{T}, \quad \Cphithree < g^{3}.
        \end{equation}
\end{itemize}
\bibliographystyle{jhep}
\bibliography{bib.bib}
\end{document}